\input epsf.tex
 \documentstyle[12pt]{article}

\newcommand{\bmat}{\left(\begin{array}}
\newcommand{\emat}{\end{array}\right)}
\def\NPB{Nucl. Phys. B}

\def\yzero{\smash{\hbox{$y\kern-4pt\raise1pt\hbox{${}^\circ$}$}}}

\def\a{\alpha}
\def\b{\beta}

\def\d{\delta}
\def\t{\times}

\def\-{\hphantom{-}}

\def\s2{\frac{1}{\sqrt2}}

\def\beq{\begin{equation}}
\def\eeq{\end{equation}}
\def\beqa{\begin{eqnarray}}
\def\eeqa{\end{eqnarray}}
\def\ba{\begin{array}}
\def\ea{\end{array}}

\def\IF{\relax{\rm I\kern-.18em F}}
\def\II{\relax{\rm I\kern-.18em I}}
\def\IP{\relax{\rm I\kern-.18em P}}
\def\IC{\relax\hbox{\kern.25em$\inbar\kern-.3em{\rm C}$}}
\def\IR{\relax{\rm I\kern-.18em R}}

\def\ti{\tilde}

\def\cp{{\cal P}}

\def\Dsl{\,\raise.15ex\hbox{/}\mkern-13.5mu D} 
\def\IZ{Z\kern-.4em  Z}

 \def\cp#1{\relax\ifmmode {\IP\kern-2pt{}_{#1}}\else $\IP\kern-2pt{}_{#1}$\=fi}

\catcode`\@=11
\newdimen\@rotdimen
\newbox\@rotbox

\def\@vspec#1{\special{ps:#1}}
\def\@rotstart#1{\@vspec{gsave currentpoint currentpoint translate
   #1 neg exch neg exch translate}}
\def\@rotfinish{\@vspec{currentpoint grestore moveto}}
\def\@rotr#1{\@rotdimen=\ht#1\advance\@rotdimen by\dp#1%
   \hbox to\@rotdimen{\hskip\ht#1\vbox to\wd#1{\@rotstart{90 rotate}%
   \box#1\vss}\hss}\@rotfinish}
\def\@rotl#1{\@rotdimen=\ht#1\advance\@rotdimen by\dp#1%
   \hbox to\@rotdimen{\vbox to\wd#1{\vskip\wd#1\@rotstart{270 rotate}%
   \box#1\vss}\hss}\@rotfinish}%
\def\@rotu#1{\@rotdimen=\ht#1\advance\@rotdimen by\dp#1%
   \hbox to\wd#1{\hskip\wd#1\vbox to\@rotdimen{\vskip\@rotdimen
   \@rotstart{-1 dup scale}\box#1\vss}\hss}\@rotfinish}%
\def\@rotf#1{\hbox to\wd#1{\hskip\wd#1\@rotstart{-1 1 scale}%
   \box#1\hss}\@rotfinish}%
\def\rotate{\@ifnextchar[{\@rotate}{\@rotate[l]}}
\def\@rotate[#1]#2{\setbox\@rotbox=\hbox{#2}\@nameuse{@rot#1}\@rotbox}

\catcode`\@=12

\begin{document}

\makeatletter
\@addtoreset{equation}{section} \makeatother
\renewcommand{\theequation}{\thesection.\arabic{equation}}
\pagestyle{empty}
\pagestyle{empty}
\rightline{FTUAM-03-30}
\rightline{IFT-UAM/CSIC-03-16}
\rightline{\today}
\vspace{0.5cm}
\setcounter{footnote}{0}

\begin{center}
{\LARGE{\bf N=1 Locally Supersymmetric Standard Models from 
Intersecting branes }}
\\[7mm]
{\Large{{  Christos ~Kokorelis} }
\\[2mm]}
\small{ \em Dep/to de F\'\i sica Te\'orica C-XI and 
Instituto de F\'\i sica 
Te\'orica C-XVI}
,\\[-0.3em]
{\em  Universidad Aut\'onoma de Madrid, Cantoblanco, 28049, Madrid, Spain}
\end{center}
\vspace{3mm}


\begin{center}
{\small \bf ABSTRACT}
\end{center}
\begin{center}
\begin{minipage}[h]{14.5cm}
We construct  
four dimensional intersecting D6-brane models 
that have locally  
the spectrum of the N=1 Supersymmetric
Standard Model. 
All open {\em visible} string sectors share the same N=1 supersymmetry.
As expected in these supersymmetric classes of models, where the D6-branes 
wrap a toroidal orientifold of type IIA,   
the hierarchy may be stabilized if the string scale 
is low, e.g. below 30 TeV. 
We analyze the breaking of supersymmetry in the vicinity of the 
supersymmetric point by turning on complex structure deformations as 
Fayet-Iliopoulos terms.
Positive masses for all squarks and sleptons, to avoid charge/colour breaking
minima, may be reached when
also two loop contributions may be included. In the ultimate version
of the present models N=1 supersymmetry may be broken by gauge mediation. 
The constructions with four, five and six stacks of D6-branes at $M_s$ are build directly. Next by the use of brane 
recombination we are able to show that 
there is a continuous, RR 
homology flow, between
six, five and four stack models. 
Moreover, we examine the gauge coupling 
constants of the Standard Model 
$SU(3)_C$,  $SU(2)_L$, $U(1)_Y$ at the string scale in the 
presence of a non-zero antisymmetric NS B-field.

\end{minipage}                 
\end{center}

\newpage
\setcounter{page}{1}
\pagestyle{plain}
\renewcommand{\thefootnote}{\arabic{footnote}}
\setcounter{footnote}{0}

\section{Introduction}

It is widely accepted that string theory provides the only
consistent theoretical framework for perturbative quantum gravity.
In this respect, in the absence of a dynamical principle for selecting 
a particular string vacuum which may be describing our world,   
one has to demonstrate beyond doubt that 
vacua with only the 
Standard Model (SM) at low energy exist. In this case, string theory may 
definitely provide us with acceptable evidence for describing the real world. 
Steps forward in this direction has been taken recently in the context of
Intersecting Brane Worlds \cite{lust1}- \cite{revo} (IBW's). We note that
string theory based intersecting brane worlds is the explicit string 
realization of the idea about the field theory 
breaking of space-time supersymmetry via the introduction of magnetic fields
\cite{bachas}. See also \cite{lustd, bia, pra}.

In general one expects that string vacua may be such that either they have 
broken supersymmetry from the start, at the string scale, or they 
will preserve some amount of 
supersymmetry \footnote{The cosmological constant problem remains 
in both options.}. 
In the latter case one might try to build explicit N=1
supersymmetric models using intersecting D6-branes, but
one has to solve first serious problems like 
the existence of extra exotic
chiral matter remaining massless to low energies - which cannot become
massive from existing stringy couplings - absence of doublet-triplet splitting for supersymmetric GUT constructions, etc.
[ See
\cite{cve1, cve, blo, ho} for studies in the context of 
N=1 SUSY intersecting 
D6 brane model building]
, before discussing realistic N=1 SUSY model building
.\newline
In the former case it has been  explicitly
demonstrated, using intersecting D6-brane language, 
that non-supersymmetric intersecting D6-brane vacua with only the SM at 
low energy exist \cite{iba, kokos1, kokos2} [For some other 
attempts to construct, using D-branes the (non-susy) SM, but not based on a particular 
string construction, see \cite{tom}.]. 
It has also been shown that non-supersymmetric GUT vacua, either
of Pati-Salam type \cite{kokos3} or SU(5)/flipped SU(5) GUTS 
\cite{kokos4} 
with only the SM at low energy exist.
In these works, it has been evident that by working, in great detail,
with four 
dimensional compactifications of type IIA toroidal orientifolds \cite{lust1}
(or their orbifolds \cite{lust3} in the case of SU(5)/flipped SU(5) 
GUTS \cite{kokos4}) -
and D6-branes intersecting at angles - one can localize eventually 
the SM at low energy in all these constructions [see for example \cite{lia1},\cite{revo} for some reviews]. We note that even though models with D6-branes 
intersecting
at angles are related by T-duality to models with magnetic deformations
 \cite{lustd, pra}, it is much easier to work in the angle picture as one may handle 
easier the parameters of the theory. 
We note that the SM at low energy has been shown to 
exist in another toroidal-like constructions, using D5-branes 
in \cite{D5}, \cite{D51}.

A number of important phenomenological issues have been also examined in 
the context of model building in IBW's,  
including proton decay \cite{igorwi, kokos4} and the implementation of 
doublet-triplet splitting mechanism \cite{kokos4}.

There is an essential difference between the four D6-brane stack models
of \cite{iba} and its five, six D6-brane stack model extensions of
 \cite{kokos1, kokos2} respectively.
 The models of \cite{iba}, which have overall N=0 supersymmetry don't have 
open string sectors which preserve some amount of supersymmetry (SUSY).
On the contrary,
the models 
of \cite{kokos1, kokos2} have necessarily some open string sectors 
which preserve N=1 supersymmetry. 
This is necessary in order to make massless the pveviously 
massive superpartners of $\nu_R$, s$\nu_R$. By giving a vev to
 s$\nu_R$ one may be able to 
break the extra U(1)'s, beyond hypercharge, that survive massless the 
Green-Schwarz anomaly cancellation mechanism. Thus at low energy only 
the SM survives. 
Similar considerations apply to the non-supersymmetric Pati-Salam GUTS of 
\cite{kokos3}, where the presence of some open string sectors
preserving N=1
supersymmetry with the orientifold plane is necessary in order to create
a Majorana mass coupling term for $\nu_R$ and get rid of the massless
beyond hypercharge U(1)'s that survive massless the Green-Schwarz 
mechanism \footnote{The existence of particular sectors is needed to 
guarantee the existence of superpartners for the right  handed neutrinos, 
s$\nu_R$.  
Thus the presense of the see-saw mechanism limits the degeneracy of the 
parameters of the RR solutions placing conditions on them. 
These conditions solve the conditions for some U(1)'s, beyond hypercharge,
to survive massless the Green-Schwarz anomaly cancellation. The latter 
U(1)'s becoming eventually massive with the help of s$\nu_R$'s.}.
In those cases, we were able to localize
the presense of the SM spectra together with extra matter at the string scale, the latter 
finally becoming massive by appropriate stringy couplings, leaving only the 
SM at low energy. \newline 
Alternatively, one may try to build non-supersymmetric models where 
each open string sector preserves some amount of SUSY, and where each brane
shares some SUSY with the other branes but not necessarily the same one.
Such constructions have been considered in \cite{iba1, iba2} and as the models are not really N=1
supersymmetric they were called quasi-supersymmetric (Q-SUSY).
The most important feature of these phenomenological models is the 
absense of one loop corrections to the scalar masses (CSM's); necessarily  
CSM's appear at two loop order.
As a result the 
lower part of the gauge hierarchy problem \footnote{Absense of 
large quadratic corrections to  the scalar masses; the higher part of 
the gauge hierarchy problem we identify as  the explanation of the 
ratio $M_W/M_{Planck}.$} 
is stabilized in these models for a string scale of less than 30 TeV. 
 The
`symmetrical' choice in the latter constructions is to choose all open 
string sectors to
share the same N=1 supersymmetry.
   In this case, we may attempt to localize the particle spectrum of the 
N=1 SUSY SM (N=1 local SUSY models), even though the models overall may 
have N=0 supersymmetry. 

In this work, firstly we consider models with intersecting D6 branes 
which share the same N=1 SUSY at every intersection. In this way we 
localize the spectrum of the N=1 
Supersymmetric Standard model. Hence we first consider 
  generalizations of the four stack 
N=1 local SM of \cite{iba2}, in the sense of allowing the most general
wrappings in the presence of NS B-field between the intersections of the branes involved. We note that 
consequences for gauge coupling unification have been examined in \cite{blulu} based
on this model. We also consider further consequences for the gauge couplings 
for the present models generalizing some of the results of \cite{blulu},
as we consider the additional presence of the NS B-field in the models of 
\cite{iba2}.   
 
Secondly, we present new solutions with the spectrum of the MSSM , 
involving a trivial NS B-field across the compact six dimensional space,
 with D-brane configurations that 
contain the five and its 
maximum extended D6-brane six 
stack constructions of \cite{iba2}.
An interesting feature of these constructions
is
that these models may have necessarily a low string scale and thus avoid  
the gauge hierarchy problem related to quadratic Higgs 
corrections \cite{iba1}. 
We also note that non-supersymmetric toroidal orientifold 
models \cite{iba1, kokos1, kokos2} or 
their orbifolds, have some non-zero NS tadpoles whose presence acts as an 
uncancelled cosmological constant, similar (but not equivalent) to the 
cosmological constant appearing after the breaking of N=1 space-time supersymmetry
in supersymmetric models.    

The paper is organized as follows. In section 2, we present a  
brief review of the main features of the constructions of \cite{iba1, iba2}.
In particular we present the most general solution to the wrapping numbers of 
the SM configurations of \cite{iba2} in the presence of a NS B-field. 
For clarity reasons we will 
identify the four, five and six stack models described, as belonging 
to the Model classes I, II and III respectively.
In sections 3, 4  we describe out new five, six stack configurations
respectively, which localize the spectrum of the N=1 SUSY SM spectrum, 
on N=1 supersymmetric intersections, in overall N=0 
models.
In section 5 we analyze the breaking of the common SUSY preserved in the
intersections using FI terms. Similar considerations have appeared in
\cite{ka, wit, cve1, iba1}. 
In section 6 we describe the existence of gauge breaking
transitions in the models appearing after brane recombination;
 the latter effectively
corresponding to the existence of adjoint Higgs fields `localized' between the
parallel recombined branes.
In section 7 we provide important formulas for the gauge coupling
constants in the models.
In section 8 we discuss gauge mediated N=1 space-time supersymmetry 
breaking and the issue of the non-factorizable brane needed to cancel RR 
tadpoles. Finally in section 9 we summarize our concluding remarks.

 
\section{N=1 local SYSY models revisited - Model I}

Our aim is to derive models which are of phenomenological 
interest and which localize the spectrum of the N=1 supersymmetric 
Standard Model (SSM).
The most important property of IBW's is that open strings stretching
in the intersection between branes get associated with 
chiral fermions localized
in their intersections \cite{dug}.
The multiplicity of the chiral fermions in the intersections is given by 
the intersection number. 
In its simplest constructions, the type IIA theory
gets compactified to four dimensions (4D) on an orientifolded $T^6$ 
torus \cite{lust1}.
Moreover, the $D6_a$ branes wrap three-cycles $(n_a^i, m_a^i)$, $i=1,2,3$, 
across each of the i-th $T^2$ torus of the assumed, for simplicity, 
factorized 
$T^6$ torus, $T^6 = T^2 \otimes  T^2 \otimes  T^2 $.  In particular, our 
torus is allowed to wrap factorized products of three 1-cycles, 
whose homology classes are given exactly by
\beq
[\Pi_a ]= \prod_{i=1}^3(n_a^i[a_i] + m_a^i [b_i])
\eeq
 and their orientifold images as 
\beq
[\Pi_{a^{*}} ]= \prod_{i=1}^3(n_a^i[a_i] - m_a^i [b_i]) \ .
\label{und1}
 \eeq
In (\ref{und1}) we have used the fact that wrapping numbers of the 
orientifold images of the $(n, m)$ wrappings of the $\a$-brane are given
by $(n, -m)$.
The wrappings numbers $(n^i, m^i)$ may take integer values in which case the tori that the 
D6-branes wrap are orthogonal. If a non-trivial NS B-field is added, the tori
becomes tilted and the effective wrapping numbers become 
$(n^i, m^i) = (n^i, {\tilde m}^i + n^i/2)$, $n^i, {\tilde m}^i,  \in Z$, where 
the {\em magnetic} wrappings $m$, take now fractional 
values. There are several open string sectors, including the $ab$ sector 
accommodating open strings stretching between the $D6_a$, $D6_b$ branes 
and localizing fermions transforming in the bifundamental representation
$(N_a, {\bar N}_b)$ with 
multiplicity
\beq
I_{ab} = [\Pi_a ]\cdot [\Pi_b ]= \prod_{i=1}^3 = (n_a^i m_b^i - m_a^i n_b^i) \ .
\eeq
Also, the $ab^{\star}$ sector is present with chiral fermions 
transforming in the
bifundamental representation $(N_a,  N_b)$ and multiplicity     
 given by 
\beq
I_{ab^{*}} = [\Pi_a ]\cdot [\Pi_{b*} ]= -\prod_{i=1}^3 (n_a^i m_b^i 
+ m_a^i n_b^i) \ .
\eeq
The sign of $I_{ab}$, $I_{ab^{*}}$, denotes the chirality
 of the corresponding
fermion and it is a matter of convention; the positive sign we 
identify with the 
left handed fermions.
 The gauge group for a set of  $a, b$ D6-branes 
on the 
above background, is in general $U(N_a) \times U(N_b)$. In the case that the brane
$a$ is its own orientifold image then the gauge group may be further 
enhanced from $U(N_a)$ to $SO(2N_a)$ or $USp(2N_a)$. The 
wrapping numbers are further constrained by the conditions set out by the 
RR tadpole cancellation conditions \cite{lust1}
\beq
\sum_a  N_a \ n_a^1 \  n_a^2 \ n_a^3  = 16,\ 
\sum_a  N_a \ n_a^1 \  m_a^2 \ m_a^3  = 0, \
\sum_a  N_a \ m_a^1 \  n_a^2 \ m_a^3  = 0, \ 
\sum_a  N_a \ m_a^1 \  m_a^2 \ n_a^3  = 0
\label{tadpo}
\eeq
\begin{itemize}
\item{N=1 Supersymmetric SM constructions from intersecting D6-branes}
\end{itemize}
Lets us assume the particular D6-brane configuration, seen in 
table (\ref{spe8}), with gauge group $U(3) \times U(1)_b  \times U(1)_c  \times U(1)_d$.
For the special values of the  
parameters $\epsilon = {\tilde \epsilon} = 1 $, $\beta_1 = \beta_2 = 1$, 
that is when all tori are untilted, we recover the D-brane configuration
proposed in \cite{iba2}. 
It clearly describes the chiral spectrum of the MSSM.
The hypercharge is defined as 
\beq
Q_Y = (1/6) \ Q_a - (1/2)\  Q_c - (1/2)\ Q_d
\label{sda1}
\eeq
Also there are exact identifications between the global symmetries of the 
SM and the U(1) symmetries set out by the D-brane configurations of table (1).
Thus baryon number ($B$) is easily identified as $Q_a = 3 B$, the lepton 
number ($L$) is given by $Q_d = L$, and $Q_c$ is twice $I_R$ the third 
component of
the right-handed weak isospin.   
\begin{table}
[htb] \footnotesize
\renewcommand{\arraystretch}{1.2}
\begin{center}
\begin{tabular}{||c|c|c|c|c|c|c||}
\hline
Matter Fields & Representation & Intersection & $Q_a$ & $Q_c$ & $Q_d$ & Y
\\\hline
 $Q_L$ &  $3(3, 2)$ & $(ab), (ab*)$ & $1$ & $0$ & $0$ &  $1/6$ \\\hline
$U_R$ & $3({\bar 3}, 1)$ & $(ac)$ & $-1$ & $1$ & $0$ &  $-2/3$ \\\hline    
 $D_R$ &   $3({\bar 3}, 1)$  &  $(a c^{\ast})$ &  $-1$ & $-1$ & $0$ &  $1/3$ \\\hline $L$ &   $3(1, 2)$  &  $({db}), (db^{\ast}) $ & $0$ & $0$ & $1$ & $-1/2$  \\\hline    
$N_R$ &   $3(1, 1)$  &  $(dc)$ &  $0$ & $1$ & $-1$ &  $0$  \\\hline    
$E_R$ &   $3(1, 1)$  &  $(dc^{\ast})$ & $0$ & $-1$ & $-1$ &  $1$  \\\hline 
$H_{d}$ & $\frac{1}{\beta_1 \cdot \beta_2}(1,2)$ & $(cb*)$ & $0$   & $ 1 $& $0$ &  $- 1/2$ \\\hline
$H_{u}$ & $\frac{1}{\beta_1 \cdot \beta_2}(1,2)$ & $ (cb)$ & $0$   & $ - 1 $& $0$ &  $ 1/2$ \\\hline
 \hline
\end{tabular}
\end{center}
\caption{\small Chiral spectrum of the four stack 
D6-brane N=1 Supersymmetric Standard Model-I with its
$U(1)$ charges. 
\label{spe8}}
\end{table}
The intersection numbers localizing the chiral fermions of table 
(\ref{spe8}) are given by 
\beq  
\begin{array}{lcl} 
I_{ab}\ =  3,  \ I_{ab*}\ = 3, I_{ac}\ = -3, \ I_{ac*}\ = -3, \ I_{bc}\ = 
-\frac{1}{\beta_1 \beta_2}, \\ 
I_{db}\ =   3, \   I_{db*}\ = 3,\ 
I_{dc}\ =  -3, \  I_{dc*}\ = -3, \ I_{bc*}\ = \frac{1}{\beta_1 \beta_2} 
\end{array} 
\label{asd12}
\eeq 
The wrapping numbers of the SM chiral fermions can be seen in table 
(\ref{spe10}). The entries of this table represent the most general solution
to the wrapping numbers (\ref{asd12}). For the special values of the 
parameters $\epsilon = {\tilde \epsilon} = 1 $, $\beta_1 = \beta_2 = 1$, 
that is when all tori are untilted, we recover the wrapping 
number solution proposed in \cite{iba2}.
Note that we have allowed for the introduction of a NS B-field, that makes the 
tori tilted along the second and the third tori. 
Lets us define \cite{iba1, kokos3} the supersymmetry vectors 
\beqa
r_1 = \pm \frac{1}{2}(----),\nonumber\\
r_2 = \pm \frac{1}{2}(-++-),\nonumber\\
r_3 = \pm \frac{1}{2}(+-+-),\nonumber\\
r_4 = \pm \frac{1}{2}(++--),
\eeqa
These vectors may help us to identify the supersymmetries shared between 
by the 
different branes of our models and the orientifold O6 plane. In general a
D6-brane, placed in a compactification of type IIA in a toroidal 
background  being a $\frac{1}{2}$ BPS state will preserve N = 4 SUSY.
If an orientifold plane is added in our configuration then there are 
non-trivial consequences for the 
supersymmetries
preserved by the D6-brane, as now the brane may have to share some 
supersymmetries with its orientifold images and subsequently the 
orientifold plane. In fact, each brane will now share a N=2 SUSY with 
its orientifold image.
 Clearly the O6 plane preserves
the SUSY described by the vector $r_1$. On the other hand the 
intersection between the
D6-branes a', b' preserves exactly the supersymmetries that are common 
between them; each brane sharing some sypersymmetries
with the orientifold. For models based on toroidal
orientifolds of type IIA \cite{lust1}, is also possible to allow for 
non-supersymmetric D6-brane 
configurations that allow each intersection to preserve either the same or a different supersymmetry with the rest of the intersections \cite{iba1, iba2}. 
This is to be contrasted with the constructions of SM's of \cite{kokos1, 
kokos2} where only the intersections where the right handed neutrino was 
localized were N=1 supersymmetric.
In the models 
we discuss in the present work the former case is realized. Thus the 
present constructions will allow the same N=1 SUSY to be equally 
shared by all intersections, achieving a fermion-boson mass degeneracy
at every intersection locally, but not globally, as globally the constructions
may be non-supersymmetric (N=0).

The Standard Models described by the 
wrappings (\ref{asd12}) have all intersections respecting the same 
N=1 supersymmetry - see table (\ref{model1}) - when the complex structure moduli \footnote{We have 
defined the complex structure as
the ratio of the radii in the corresponding tori, namely 
$\chi_i = \frac{R_2^{(i)}}{R_1^{(i)}}$. The value $\beta_i =1/2$ signals the 
presence of a non-zero B-field \cite{beta}. 
} satisfy
\beq
\b_1 \cdot \chi_2 = \b_2 \cdot \chi_3 \ ,
\label{li1}
\eeq
where we have defined 
\beq
\a_1 = tan^{-1}(3 \rho^2 \b_1 \chi_1)
\eeq
Let us now focus our attention to the gauge symmetry of the models. After the
  implementation of the Green-Schwarz anomaly cancellation mechanism, 
the actual gauge group of the N=1 local SUSY Standard Model I
\begin{table}
[htb]\footnotesize
\renewcommand{\arraystretch}{1.5}
\begin{center}
\begin{tabular}{||c||c|c|c||} 
\hline
\hline
$\ N_i$ & $(n_i^1, \ m_i^1)$ & $(n_i^2, \ m_i^2)$ & $(n_i^3, \ m_i^3)$\\
\hline\hline
 $N_a=3$ & $(1,\ 0)$  &
$(1/\rho, \  3 \rho \epsilon \beta_1)$ & $(1/\rho, \ -3\rho {\tilde \epsilon} 
\beta_2)$  \\
\hline
$N_b=1$  & $(0, \ \epsilon {\tilde \epsilon})$ & $(1/\beta_1,\ 0)$ & 
$(0, \ -{\tilde \epsilon})$ \\
\hline
$N_c=1$ & $(0, \ \epsilon)$ & $(0,\  -\epsilon)$  & 
$({\tilde \epsilon}/\beta_2, \ 0)$ \\    
\hline
$N_d=1$ & $(1,\ 0)$ &  $(1/\rho,\  3 \rho \epsilon \beta_1 )$  
  & $(1/\rho, \ -3 \rho {\tilde \epsilon} \beta_2)$  \\\hline
\end{tabular}
\end{center}
\caption{\small General wrapping numbers of Model I, 
giving rise to the N=1
Standard Model. The wrappings depend on the parameters $\rho = 1, 1/3$; 
$\epsilon = {\tilde \epsilon}= \pm 1$ and the NS background on the last two 
tori $\b_i = 1 - b^i = 1, 1/2$. Hence they describe eight different 
sets of wrappings giving rise to the same intersection numbers.
\label{spe10}}          
\end{table}
becomes $SU(3) \times SU(2)_L \times U(1)_Y \times U(1)_X$, where 
\beq
U(1)_X =\ 3 Q_a -\ 9 Q_d +\ 10 Q_c
\eeq
and we have 
assumed that the brane $b$ has been brought on top of its orientifold 
image $b^{\star}$. The extra $U(1)_X$ generator may be broken by giving a vev to the
right handed neutrino $s\nu_R$. 
A comment is in order at this point. 
We note that the initial gauge symmetry of the models may be further 
enhanced \footnote{as have been already noted in \cite{iba2}}
to the Pati-Salam 
\beq
 SU(4) \times SU(2)_L \times 
SU(2)_R
\label{pati}
\eeq if the brane $d$ is brought on top of brane $a$ and
the brane $c$ is brought on top of its orientifold image. In this case,
breaking the Pati-Salam symmetry $SU(4) \rightarrow SU(3) \times U(1)$ 
corresponds to moving apart the branes $a, d$ 
along different points of the complex plane; 
giving vev's to the adjoint multiplets localized in the intersection between
the branes. 
We should also note that when the gauge symmetry of the theory is in the form
of a GUT group, like of the  
Pati-Salam type (\ref{pati}), and one necessarily has to 
have some sectors preserving a SUSY, 
  one should expect some of the branes 
$a, b, c, d$ to form
angles $\pi/4$ with respect to the orientifold plane. 
This have been firstly observed
in all the non-supersymmetric toroidal Pati-Salam orientifold 
constructions of 
\cite{kokos3} with only the SM at low energy - where on phenomenological grounds some sectors were N=1 
SUSY preserving - and also observed lately in a Pati-Salam N=1 SUSY GUT example in \cite{ho}.

Assuming the moduli fixing relation (\ref{li1}) the model shares in each intersection the same N=1 supersymmetry with the orientifold plane. 
On the other hand,
as have been also discussed in \cite{ka, wit} adjusting 
the complex structure moduli slightly off their SUSY values results
in the breaking of supersymmetry [These issues will be discussed in detail
in section 6]. 
The wrapping numbers corresponding to the D6-brane configuration of table (1),
do not satisfy RR tadpoles. In general RR tadpoles may be satisfied in a
number of ways. In the present models the D6-brane configuration of 
table (1) is free of any gauge and mixed U(1)-gauge anomalies.
Thus RR tadpoles may be 
cancelled by adding a non-factorizable (NF) D6 brane which 
has intersections with the SM 'visible' branes but where
the produced chiral fermions may become massive
from existing stringy couplings and simultaneously 
don't contribute to anomalies. In the latter case this 
messenger sector may contain $N_f$ flavours of chiral superfields 
$\Phi_I$, ${\bar \Phi}_i$ transforming in the 
representations ${\bar {\bf r}} + {\bar{\bf r}}$ of the 
messenger gauge group. 
Hence the combined 
system of the D6-brane/NF-brane may be non-supersymmetric
and the messenger sector may be responsible for breaking the supersymmetry 
through
gauge mediation \cite{dine}.
The same conclusion may be reached
for local N=1 SM's II and III. These issues will be further analyzed
in section 8.

\begin{table}
\renewcommand{\arraystretch}{0.8}
\begin{center}
\begin{tabular}{||c||c|c|c|c||}
\hline
\hline
$Brane$ & $\theta_a^1$ & $\theta_a^2$ & $\theta_a^3$ & $SUSY preserved$\\
\hline\hline
 $a$ & $0$  &
$\a_1$ & $-\a_1$ & $r_2, r_1$\\
\hline
$b$  & $\frac{\pi}{2}$ & $0$ &
$-\frac{\pi}{2}$  & $ r_3, r_1$ \\
\hline
$c$ & $\frac{\pi}{2}$ &
$-\frac{\pi}{2}$  & $0$ &  $r_4, r_1$\\    
\hline
$d$ & $0$ &  $ \a_1$  
  & $ -\a_1$  & $r_2, r_1$\\\hline
\hline
\end{tabular}
\end{center}
\caption{\small 
The N=2 supersymmetries shared by the different intersections for the
four stack N=1 Supersymmetric Standard Model I.  
\label{model1}}          
\end{table}

\begin{itemize}

\item{ The N=1 Higgs system}

\end{itemize}

The number of Higgses present depends on the number of tilted tori in the 
models. Their charges may be read easily from table (\ref{spe8}).
At the level of effective theory  these Higgses will appear as a 
mixture of the fields $H^{+} =  H_u + H_d^{*}$,  $H^{-} =  H_d + H_u^{*}$.
Thus we can have either one pair of Higgs fields 
with $\b_1 = \b_2 =1$ in 
which case the chiral fermions of table (1) represent the spectrum of the MSSM, or 
 $\b_1 =1/2,  \b_2 =1/2$  which is doubling the 
Higgs content of MSSM. 
Another choice will be when
  $\b_1 = \b_2 =1/2$ (or $\b_1 =1/2,  \b_2 =1$), where only the one of the 
Higgs doublet of the MSSM is doubled.


\section{N=1 local SM vacua from five stacks of Intersecting D6-branes - Model II}


In this section, we will analyze some five stack D6-brane models 
which are constructed in such a way that the charged lepton and neutrino
generations are localized in different intersections. As it will be 
explained later this has non-trivial 
consequences for the Yukawa couplings of the models. Our main aim in this 
section is to exhibit the basic properties of these D6-brane configurations.  

\subsection{Supersymmetric Standard Model II} 
Our D6-brane 
configuration may be seen in table (\ref{spe82}). 
The initial generic gauge symmetry of the models is 
\beq
U(3)_a \times 
U(1)_b  \t U(1)_c \t U(1)_d \t  U(1)_e\ ,
\eeq
further enhanced  - when the brane b is brought on top of 
its orientifold image - to
\beq
U(3)_a \times 
SP(2)_b  \t U(1)_c \t U(1)_d \t  U(1)_e \    .
\label{assu1}
\eeq
The intersection numbers giving rise to the chiral spectrum of the SM 
in table (\ref{spe82}) may   
\begin{table}[htb] \footnotesize
\renewcommand{\arraystretch}{1.2}
\begin{center}
\begin{tabular}{||c|c|c|c|c|c|c|c||}
\hline
Matter Fields & Representation & Intersection & $Q_a$ & $Q_c$ & $Q_d$ & $Q_e$& Y
\\\hline
 $Q_L$ &  $3(3, 2)$ & $(ab), (ab*)$ & $1$ & $0$ & $0$ & $0$ & $1/6$ \\\hline
$U_R$ & $3({\bar 3}, 1)$ & $(ac)$ & $-1$ & $1$ & $0$ & $0$ & $-2/3$ \\\hline    
 $D_R$ &   $3({\bar 3}, 1)$  &  $(a c^{\ast})$ &  $-1$ & $-1$ & $0$ & $0$ & $1/3$ \\\hline $L$ &   $2(1, 2)$  &  $({db}), (db^{\ast}) $ & $0$ & $0$ & $1$ & $0$ & $-1/2$  \\\hline    
$l_L$ &   $(1, 2)$  &  $(be), (be^{\ast}) $ & $0$ & $0$ & $0$ & $1$ & $-1/2$  \\\hline    
$N_R$ &   $2(1, 1)$  &  $(dc)$ &  $0$ & $1$ & $-1$ & $0$ & $0$  \\\hline    
$E_R$ &   $2(1, 1)$  &  $(dc^{\ast})$ & $0$ & $-1$ & $-1$ & $0$ & $1$  \\\hline
  $\nu_R$ &   $(1, 1)$  &  $(c e)$ &  $0$ & $1$ & $0$ & $-1$ & $0$\\\hline
$e_R$ &   $(1, 1)$  &  $(c e^{\ast})$ &  $0$ & $-1$ & $0$ & $-1$  & $1$ 
\\\hline  
$H_{d}$ & $\frac{1}{\beta_1 \cdot \beta_2}(1,2)$& $(cb*)$ & $0$   & $  1 $& $0$ & $0$ & $- 1/2$ \\\hline
$H_{u}$ & $\frac{1}{\beta_1 \cdot \beta_2}(1,2)$& $ (cb)$ & $0$   & $ - 1 $& $0$ & $0$ & $ 1/2$ \\\hline
 \hline
\end{tabular}
\end{center}
\caption{\small Chiral spectrum of the five stack 
D6-brane Model-II with its
$U(1)$ charges. The hypercharge is defined as $Q_Y = \frac{1}{6} Q_a -
 \frac{1}{2} Q_c -\frac{1}{2} Q_d -\frac{1}{2} Q_e $.
\label{spe82}}
\end{table}
be described by 
\beq 
\begin{array}{lcl} 
I_{ab}\ =   \ 3, &  I_{ab*}\ =\ 3, \ \ I_{ac}\ =   \ -3,  &\  I_{ac*}\ =\ -3, \\ 
I_{db}\ =   \ 2,  &   I_{db*}\ =\ 2, \ \  I_{be}\ =   \ -1,  & \ I_{be*}\ =\ 1, \\ 
I_{dc}\ =   \ -2, &  I_{dc*}\ =\ -2,\ \ I_{ce}\ =   \ 1,  & \ I_{ce*}\ =\ -1, \\
I_{bc}\ =   \ - \frac{1}{\beta_1 \beta_2}, & I_{bc*}\ =\ \frac{1}{\beta_1 \beta_2},& 
\end{array} 
\eeq 
The D6-brane wrappings satisfying these intersection numbers are 
given in table (\ref{spe1010}). In the present SM's II
 the global symmetries of the SM get
 identified as
\beq
Q_a = 3 B, \ L = Q_d + Q_e, \ Q_c = 2 I_R
\eeq 
\begin{table}
\renewcommand{\arraystretch}{1.5}
\begin{center}
\begin{tabular}{||c||c|c|c||}
\hline
\hline
$\ N_i$ & $(n_i^1, \ m_i^1)$ & $(n_i^2,\  m_i^2)$ & $(n_i^3,\ m_i^3)$\\
\hline\hline
 $N_a=3$ & $(1, 0)$  &
$(1/\rho, \ 3 \rho \epsilon \beta_1)$ & $(1/\rho, \ -3 \rho {\tilde \epsilon}
 \beta_2) $ \\
\hline
$N_b=1$  & $(0, \ \epsilon {\tilde \epsilon} )$ & $(1/\beta_1,\  0)$ & $(0,\ -
{\tilde \epsilon})$ \\
\hline
$N_c=1$ & $(0,\ \epsilon )$ &
$(0, \ -\epsilon)$  & $({\tilde \epsilon}/\beta_2,\  0)$ \\    
\hline
$N_d=1$ & $(1, \ 0)$ &  $(1, \  2 {\epsilon} \beta_1)$  
  & $(1, \ -2 {\tilde \epsilon} \beta_2)$  \\\hline
$N_e = 1$ & $(1, \ 0)$ &  $(1,\ {\epsilon} \beta_1 )$  
  & $(1,\ -{\tilde \epsilon} \beta_2)$  \\
\hline
\end{tabular}
\end{center}
\caption{\small Wrapping numbers of Model II, 
giving rise to the N=1 Supersymmetric 
Standard Model. These wrappings depend on the phase parameters 
$\epsilon = {\tilde \epsilon} =  \pm 1$, the parameters $\b_1 =\b_2=1, 1/2$
parametrizing the NS B-field and the parameter $\rho = 1, 1/3$.
\label{spe1010}}          
\end{table}
One can easily confirm that if the condition (\ref{li1})
holds then the brane configuration respects the same N=1 SUSY at every 
intersection.
In fact, the supersymmetries preserved by the brane content of the models may be found
in table (\ref{rou111}). The angle structure of the branes with respect to the 
orientifold planes may be seen in table (\ref{rou111}), 
where we have defined
\beq 
\alpha_2 = tan^{-1}(\b_1 \chi_2), \ \gamma_1 = tan^{-1}( 2 \b_1 \chi_2)
\label{def} 
\eeq
and 
\beq
F_L = (4, 2, 1), \ \ {\bar F}_R = (4, 1, {\bar 2})
\eeq
the matter multiplets (see \cite{kokos3} for detailed realizations of 
the 
Pati-Salam GUT in intersecting brane worlds and classes of models with
only the SM at low energy) of the Pati-Salam 
$SU(4)_C \t SU(2)_L \t SU(2)_R$ GUT.

 By implementing the Green-Schwarz anomaly 
cancellation mechanism  \cite{iba} and 
analyzing the U(1) BF couplings to the RR fields 
\beq
N_a \cdot m_a^1 \cdot m_a^2 \cdot m_a^3 \cdot \int_{M_4}\ B_2^0 \wedge F_a,\ 
N_a \cdot m_a^I \cdot m_a^J \cdot m_a^K \cdot \int_{M_4}\ B_2^I \wedge F_a,
\eeq
we find, the rest of the couplings having a zero stregth with the RR fields,
\beqa
B_2^1 \wedge (\epsilon \b_1)\cdot (9 F^a + 2 F^d + F^e),\nonumber\\
B_2^3 \wedge (-{\tilde \epsilon} \b_2)\cdot (9 F^a + 2 F^d + F^e);
\eeqa
and thus conclude that only one anomalous U(1),
$9 Q_a + 2 Q_d + Q_e$ gets massive by having a non-zero coupling to RR fields,
 the other three U(1)'s 
\beqa
Q^{(1)} = \frac{1}{6}(Q_a - 3 Q_d - 3 Q_e) -\frac{1}{2}Q_c,\
Q^{(2)} = \frac{1}{6}(Q_a - 3 Q_d - 3 Q_e) +\frac{19}{18}Q_c,\\
Q^{(3)} = (-\frac{3}{28}Q_a + Q_d -\frac{29}{28}Q_e) 
\label{rota1}
\eeqa
remain massless [See \cite{anoma} for further consequences of  
for gauge field anomalies in general
orientifold models .].
The $Q^{(1)}$ generator is identified as the hypercharge in the models.
We note that at this point the actual gauge group of the models is 
$SU(3) \times SU(2) \times U(1)_Y \times Q^{(2)} \times Q^{(3)}$.  
The 
additional U(1)'s $Q^{(2}$, $Q^{(3)}$ that survive massless the Green-Schwarz
anomaly cancellation mechanism, can be broken by giving vev's to the right
handed sneutrinos, s$N_R$'s,
 localized in the $dc$ intersection. Alternatively one may choose 
to break them by giving vev's to two linear combinations of s$N_R$'s, s$\nu_R$'s from
neutrinos localized in the intersections $dc, ce$ respectively. 
This mechanism may be 
described by the presence of the Fayet-Iliopoulos terms for the anomalous 
U(1)'s and will be described later on. 

\subsection{Gauge Group Enhancement} 

The initial gauge symmetry of the models, before implementation of 
Green-Schwarz mechanism,  may be further enhanced in a number of 
different ways.
By choosing for example
the values $\rho =  1/3$, $\epsilon = {\tilde \epsilon} $ in table 5, it 
is clear that
\begin{figure}
\centering
\epsfxsize=6in
\hspace*{0in}\vspace*{.0in}
\epsffile{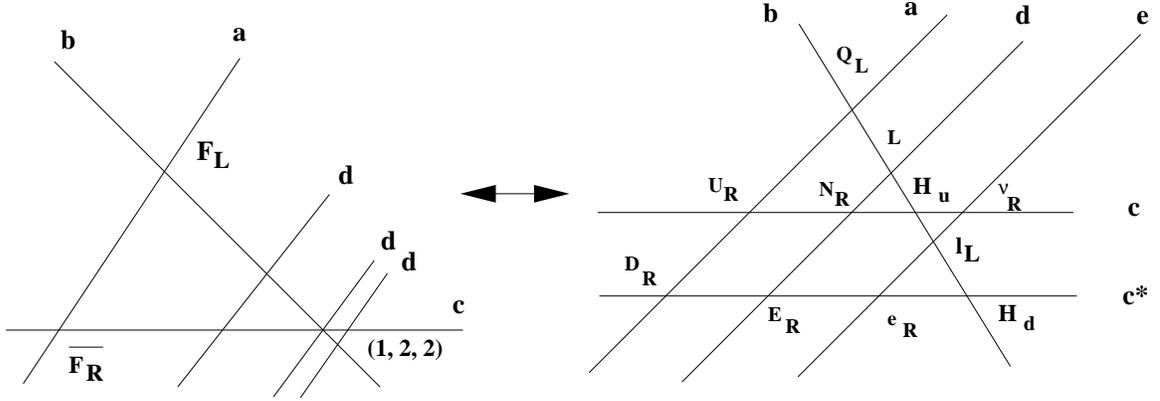}
\caption{\small Gauge group enhancement (and vice-versa) in the five stack quiver to
 a Pati-Salam $SU(4) \times SU(2)_L \times SU(2)_R \times U(1)^d$. The brane
d on the left is shown to different positions.
}
\end{figure}
as the branes $a, e$ are parallel, 
by placing brane $a$ on top of brane $e$, we may have a
symmetry enhancement from 
\beq
U(3)_a \times U(1)_e \rightarrow U(4)
\label{ska1}
\eeq
Also, if branes b, c are brought on top of their
\begin{table}
[htb]\footnotesize
\renewcommand{\arraystretch}{0.8}
\begin{center}
\begin{tabular}{||c||c|c|c|c||}
\hline
\hline
$Brane$ & $\theta_a^1$ & $\theta_a^2$ & $\theta_a^3$ & $SUSY preserved$\\
\hline\hline
 $a$ & $0$  &
$\a_1$ & $-\a_1$ & $r_2, r_1$\\
\hline
$b$  & $\frac{\pi}{2}$ & $0$ &
$-\frac{\pi}{2}$  & $ r_3, r_1$ \\
\hline
$c$ & $\frac{\pi}{2}$ &
$-\frac{\pi}{2}$  & $0$ &  $r_4, r_1$\\    
\hline
$d$ & $0$ &  $ \gamma_1$  
  & $ -\gamma_1$  & $r_2, r_1$\\\hline
$e$ & $0$ &  $\a_2$  
  & $-\a_2$  & $r_2, r_1$\\
\hline
\end{tabular}
\end{center}
\caption{\small 
The N=2 supersymmetries shared by the different intersections for the
five stack pentagonal quiver minimal SUSY Model II of table (\ref{spe82}).  
\label{rou111}}          
\end{table}
orientifold images, then the full 
gauge symmetry of the models 
may be further enhanced to 
\beq U(4) \times SU(2)_L^b \times SU(2)_R^c 
\times U(1)_d
\label{ska2}
\eeq


\section{The N=1 Standard Model from Six-stacks of Intersecting D6-branes - Model III}

In this section, we will discuss the extension of the four and five stack 
supersymmetric Standard Models I, II to their closest generalizations
with six stacks of D6-branes at the string scale. These models constitute the 
maximum allowed deformations as it may easily seen from the D-brane
configurations of tables (1) and (4). The initial gauge group structure 
is an
\beq
U(3) \t U(1)_b \t U(1)_c \t U(1)_d \t U(1)_e \t U(1)_f 
\eeq
further enhanced to 
\beq
U(3) \t SP(2) \t U(1)_c \t U(1)_d \t U(1)_e \t U(1)_f 
\label{da12}
\eeq 
when brane b is brought on top of its orientifold image.   
 
\begin{table}[htb] \footnotesize
\begin{center}
\begin{tabular}{||c|c|c|c|c|c|c|c|c||}
\hline
Matter Fields & Representation & Intersection & $Q_a$ & $Q_c$ & $Q_d$ 
& $Q_e$&  $Q_f$  & Y
\\\hline
 $Q_L$ &  $3(3, 2)$ & $(ab), (ab*)$ & $1$ & $0$ & $0$ & $0$ & $0$  &$1/6$ \\\hline
$U_R$ & $3({\bar 3}, 1)$ & $(ac)$ & $-1$ & $1$ & $0$ & $0$ & $0$ & $-2/3$ \\\hline    
 $D_R$ &   $3({\bar 3}, 1)$  &  $(a c^{\ast})$ &  $-1$ & $-1$ & $0$ & $0$ 
&  $ 0 $    & $1/3$ \\\hline 
$L^1$ &   $(1, 2)$  &  $({db}), (db^{\ast}) $ & $0$ & $0$ & $1$ & $0$ & $0$  &$-1/2$  \\\hline    
$L^2$ &   $(1, 2)$  &  $(be), (be^{\ast}) $ & $0$ & $0$ & $0$ & $1$ &$0$ &$-1/2$  \\\hline 
$L^3$ &   $(1, 2)$  &  $(bf), (bf^{\ast}) $ & $0$ & $0$ & $0$ & $0$ & $1$& $-1/2$  \\\hline 
$N_R^1$ &  $(1, 1)$  &  $(cd)$ &  $0$ & $1$ & $-1$ & $0$ & $0$ & $0$  
\\\hline    
$E_R^1$ &   $(1, 1)$  &  $(cd^{\ast})$ & $0$ & $-1$ & $-1$ & $0$  & $1$ 
& $1$ \\\hline
 $N_R^2$ &  $(1, 1)$  &  $(ce)$ &  $0$ & $1$ & $0$ & $-1$ & $0$ & $0$  
\\\hline    
$E_R^2$ &   $(1, 1)$  &  $(ce^{\ast})$ & $0$ & $-1$ & $0$ & $-1$  & $0$ 
& $1$ \\\hline
$N_R^3$ &  $(1, 1)$  &  $(cf)$ &  $0$ & $1$ & $0$ & $0$ & $-1$ & $0$  
\\\hline    
$E_R^3$ &   $(1, 1)$  &  $(cf^{\ast})$ & $0$ & $-1$ & $0$ & $0$  & $-1$ 
& $1$ \\\hline
$H_{u}$ & $\frac{1}{\beta_1 \cdot \beta_2}(1,2)$& $(cb*)$ & $0$   & $  1 $& $0$ & $0$ & $0$ &$- 1/2$ \\\hline
$H_{d}$ & $\frac{1}{\beta_1 \cdot \beta_2}(1,2)$& $(cb)$ & $0$   & $ - 1 $& $0$ & $0$ & $0$ &$ 1/2$ \\\hline
 \hline
\end{tabular}
\end{center}
\caption{\small Standard model chiral spectrum of the six stack 
string scale D6-brane N=1 SUSY Model III together with its
$U(1)$ charges. 
\label{spec811}}
\end{table}

The chiral spectrum of the SUSY SM gets localized at the various intersections 
as described in the D-brane configuration of table (\ref{spec811}).
The observed chiral spectrum gets reproduced by the following intersection 
numbers
\beq 
\begin{array}{rcc} 
I_{ab}\ =    3, &  I_{ab*}\ =\ 3,\  I_{ac}\ =   \ -3,  & I_{ac*}\ =\ -3, \\ 
I_{db}\ =  1,  & I_{db*}\ =\ 1, \  I_{eb}\ =  \ 1,  & I_{eb*}\ =\ 1, \\
I_{fb}\ =  1, &  I_{fb*}\ =\ 1,\  I_{dc}\ =   \ -1,  & I_{dc*}\ =\ -1,\\
I_{ec}\ =    -1, &  I_{ec*}\ =\ -1, \ I_{fc}\ =   \ -1,  & I_{fc*}\ =\ -1, \\
I_{bc}\ =   \ -\frac{1}{\beta_1 \cdot \beta_2}, & & I_{bc*}\ =\ \frac{1}{\beta_1 \cdot \beta_2}, \\
\end{array} 
\eeq 
where all other intersections are vanishing.
\begin{table}[htb]\footnotesize
\renewcommand{\arraystretch}{1.5}
\begin{center}
\begin{tabular}{||c||c|c|c||}
\hline
\hline
$\ N_i$ & $(n_i^1, \ m_i^1)$ & $(n_i^2,  \ m_i^2)$ & $(n_i^3, \ m_i^3)$\\
\hline\hline
 $N_a=3$ & $(1,\  0)$  &
$(1/\rho, \ 3 \rho \epsilon \beta_1)$ & $(1/\rho, \ -3 \rho {\tilde \epsilon}
 \beta_2) $ \\
\hline
$N_b=2$  & $(0, \ \epsilon {\tilde \epsilon} )$ & $(1/\beta_1,\  0)$ & $(0,\  -
{\tilde \epsilon})$ \\
\hline
$N_c=1$ & $(0, \ \epsilon )$ &
$(0, \ -\epsilon)$  & $({\tilde \epsilon}/\beta_2, \ 0)$ \\    
\hline
$N_d=1$ & $(1,\ 0)$ &  $(1,  \ {\epsilon} \beta_1)$  
  & $(1, \ - {\tilde \epsilon} \beta_2)$  \\\hline
$N_e = 1$ & $(1,\  0)$ &  $(1,\ {\epsilon} \beta_1 )$  
  & $(1,\  -{\tilde \epsilon} \beta_2)$  \\\hline
$N_f = 1$ & $(1,\ 0)$ &  $(1,\ {\epsilon} \beta_1 )$  
  & $(1, \ -{\tilde \epsilon} \beta_2)$ \\\hline
\hline
\end{tabular}
\end{center}
\caption{\small Wrapping numbers of Model III, 
giving rise to the N=1
standard model. The parameter $\rho = 1, 1/3$ describes two different sets of wrappings giving rise to the same intersection numbers.
\label{sp101}}          
\end{table}
The global symmetries of the SM may be expressed in terms of the U(1)
symmetries of the models as 
\beq
Q_a = 3 B, \ \ L = Q_d + Q_e + Q_f, \ \  Q_c = 2 I_R
\eeq 
Analyzing the U(1) BF couplings to the RR fields we conclude that
the anomalous U(1),
$9 Q_a +  Q_d + Q_e + Q_f$ gets massive by having a non-zero coupling 
to RR fields, and other four U(1)'s remain massless, the following 
\beqa
{\tilde Q}^{(Y)}& =& \frac{1}{6}(Q_a - 3 Q_d - 3 Q_e -3 Q_f) -\frac{1}{2}Q_c,\\
{\tilde Q}^{(1)}& =& \frac{1}{6}(Q_a - 3 Q_d - 3 Q_e -3 Q_f) +\frac{28}{18}Q_c,\\
{\tilde Q}^{(2)} &=& (-2 Q_d + Q_e + Q_f),\\
{\tilde Q}^{(3)} &=& (Q_e - Q_f) \ .
\label{rota11}
\eeqa
The ${\tilde Q}^{(Y)}$ is identified as the hypercharge in the models.
The breaking of the remaining U(1)'s proceeds once the sneutrino's, s$N_R$'s
get a vev. An obvious choice will be for 
${\tilde Q}^{(1)}$ to be broken by a vev from s$N_R^1$, 
${\tilde Q}^{(2)}$ to be broken by a vev from s$N_R^2$, and
 ${\tilde Q}^{(3)}$ to be broken by a vev from s$N_R^3$. Appropriate 
combinations of vev's by s$N_R$'s can also have the same effect on the
Higgsing of the U(1)'s.
The SUSY's shared by each intersection with the orientifold plane 
may be seen in table (\ref{susix}). 

\begin{table}
\renewcommand{\arraystretch}{1}
\begin{center}
\begin{tabular}{||c||c|c|c|c||}
\hline
\hline
$Brane$ & $\theta_a^1$ & $\theta_a^2$ & $\theta_a^3$ & $SUSY preserved$\\
\hline\hline
 $N_a=3$ & $0$  & $\a_1$ & $-\a_1$ & $r_2, r_1$\\
\hline
$N_b=2$  & $\frac{\pi}{2}$ & $0$ & $-\frac{\pi}{2}$  & $ r_3, r_1$ \\
\hline
$N_c=1$ & $\frac{\pi}{2}$ & $-\frac{\pi}{2}$  & $0$ &  $r_3, r_4$\\    
\hline
$N_d=1$ & $0$ &  $ \a_2$  & $ -\a_2$  & $r_2, r_1$\\\hline
$N_e = 1$ & $0$ &  $\a_2$  & $-\a_2$  & $r_2, r_1$\\\hline
$N_f = 1$ & $0$ &  $\a_2$  & $-\a_2$  & $r_2, r_1$\\
\hline
\end{tabular}
\end{center}
\caption{\small 
The N=2 supersymmetries shared by the different intersections for the
six stack N=1 SUSY Model-III.  
\label{susix}}          
\end{table}

The gauge symmetry may be enhanced to SU(5) when (we choose $\rho =1/3$) 
the branes d, e are brought on top of brane a; with
 $SU(3) \t U(1)_d \t U(1)_e 
\rightarrow   SU(5)$. In this case the full gauge group becomes 
\beq
SU(5) \t U(1)_b \t U(1)_c \t U(1)_f
\eeq or even further enhanced to 
\beq
SU(6) \t U(1)_b \t U(1)_c 
\eeq
if brane f is brought on top of brane a.
 Alternatively, we could choose
to further enhance the gauge group by bringing together the b, c branes
on top of their orientifold images. The final gauge group 
in this case will be \footnote{When $\b_1 = \b_2$ one may locate the
brane f on top of its orientifold image, extending the gauge group
to  $SU(5) \t SU(2)_L \t SU(2)_R \t SU(2)_f$.}
\beq 
SU(5) \t SU(2)_L \t SU(2)_R \t U(1)_f
\eeq
or if brane f is located on top of brane a to
 \beq 
SU(6) \t SU(2)_L \t SU(2)_R 
\eeq
If brane b (or alternatively brane c) is not brought on top of its respective orientifold image $b^{*}$,
then the gauge group may be $SU(5) \t U(1)_b \t SU(2)_R \t U(1)_f$ (
$SU(5) \t SU(2)_L \t U(1)_c \t U(1)_f$).  
The reverse procedure of the gauge group enhancement that is the actual
splitting of i.e. the $SU(5) \rightarrow SU(3) \t U(1) \t U(1)$ corresponds
to adjoint breaking.

\subsection{Yukawa Couplings }

The tree level Yukawa couplings for the 
N=1 
SM's I, II III
have several differences as the charged lepton and neutrino species are 
localized to different intersection points. 
The Higgs fields in all models are localized between the b,c and b, c*,
 branes and are shown in table (\ref{Higgs}). We assume the existence of the
D-brane configuration of tables (1), (4), (7).
 \begin{table} [htb] \footnotesize
\renewcommand{\arraystretch}{1}
\begin{center}
\begin{tabular}{||c|c|c|c||}
\hline
\hline
Intersection & EW Higgs & $\ Q_c$ & $Y$\\
\hline\hline
$\{ bc \}$ & $h_1$  & $-1$ & $1/2$\\
\hline
$\{ bc \}$ & $h_2$  & $1$& $-1/2$ \\
\hline
$\{ bc* \}$  & $H_1$ & $-1$ &  $1/2$\\
\hline
$\{ bc* \}$  & $H_2$ & $1$  &$-1/2$\\
\hline
\end{tabular}
\end{center}
\caption{\small Higgs fields responsible for electroweak symmetry breaking in
the N=1 SUSY type I, II and III models.
\label{Higgs}}
\end{table}

The tree level Yukawa couplings, that are being allowed by gauge and charge conservation invariance, are 
common for the quark sector to models I, 
II and III - and they are given by
\beq
{\cal L}^I \propto \  \ h_U^j Q_L U_R^j h_1 \ +  \  h_D^j Q_L D_R^j H_2 \ 
\eeq
Also the tree level Yukawa's for the charged lepton and neutrino 
sectors of models I, II, III, are given respectively by
\beqa
{\cal L}^I \propto \ \  y_L^{ij} L^i E_R^j H_2 \ + \ Y_N^{ij} L^i N_R^j h_1\ ,
\nonumber\\
{\cal L}^{II} \propto \ y_L^{i} L^i E_R^i H_2 \ + \ y_e l_L e_R h_2 \ 
+ \ y_{N_R} L N_R h_1 \ + \ 
 y_{\nu_R}l_L \nu_R h_1 \ ,               \nonumber\\
{\cal L}^{III} \propto \ ( \sum_{k=1}^3 L^k E_R^k h_2)\ +\ ( \sum_{k=1}^3 L^k N_R^k h_1)
\eeqa
where i=1,2; j =1,2,3.




\section{Supersymmetry breaking and sparticle masses}

As have been noted earlier for special values of the complex structure $\chi$
moduli all intersections in the models may share the same 
N=1 SUSY with the orientifold plane, that is the same N=1 SUSY is preserved at each
intersection.  In general, as has been noted in \cite{wit} 
where configurations  
with non-zero B-field and two D-branes 
each one preserving N=1 supersymmetry were used, by turning on a 
Fayet-Iliopoulos (FI) term one may able to break  
supersymmetry.
These FI terms may be computed by the use of 
the supersymmetric completion of the Chern-Simon 
\beq
\sim \sum_{i=1}^{b_3}\sum_{a=1}^{l}\int d^4x \  Z_{ij} \ B_i \wedge F_a  
\eeq
couplings \cite{sei} involved in the Green-Schwarz anomaly cancellation 
mechanism
appearing with the usual auxiliary \footnote{By $\phi_a$ we denote the
superpartners of the Hodge duals of the RR fields; $ K$ is the K\"ahler 
potential.}
 field $D_a$.
\beq
\sim \sum_{i=1}^{b_3}\sum_{a=1}^{l}\int d^4x \  Z_{ij} \frac{\partial K}
{\partial \phi} D_a
\eeq 
For an effective gauge theory involving D-branes
the presence of a FI term is then interpreted 
as coupling the complex structure moduli as FI-terms in the effective 
theory \cite{ka, wit}. Assuming a small departure of the complex structure from the 
supersymmetric situation - the SUSY wall -  the
FI term 
\beq
V_{FI} = \frac{1}{2 g_{\a}^2}(\sum_{i} q^i_{\a} |\phi^i| +\xi )^2
\eeq
captures the leading order effect to the mass 
of the scalar - we assume the existence of two branes  $D6_{\sigma}$, $D6_{\tau}$  
\beq
\alpha^{\prime}m^2_{\sigma \tau} = - q_{\sigma}  \xi_{\sigma} - 
q_{\tau} \xi_{\tau}
\eeq
leaving at an intersection $\sigma \tau$ -   as it is 
described from string theory \cite{cve1}. Related issues have also been 
examined 
in \cite{iba1,iba2}. 
In this section, we will make use of FI terms
in order to break N=1 SUSY at the various intersections of the N=1 models I, II, III. 
 \newline \newline
{\bf N=1 SUSY Model I $\&$ II}
\begin{table}
[htb]\footnotesize
\renewcommand{\arraystretch}{0.8}
\begin{center}
\begin{tabular}{||c||c|c|c|c||}
\hline
\hline
$Brane$ & $\theta_a^1$ & $\theta_a^2$ & $\theta_a^3$ & $approx.\  SUSY \ preserved$\\
\hline\hline
 $a $ & $0$  &
$\a_1 + \delta_a$ & $-\a_1$ & $r_1, r_4$\\
\hline
$b $  & $\frac{\pi}{2} + \delta_b$ & $0$ &
$-\frac{\pi}{2}$  & $ r_2, r_4$ \\
\hline
$c $ & $\frac{\pi}{2} + \delta_c$ &
$-\frac{\pi}{2}$  & $0$ &  $r_3, r_4$\\    
\hline
$d $ & $0$ &  $ \gamma_1 + \delta_d$  
  & $ -\gamma_1$  & $r_1, r_4$\\\hline
$e $ & $0$ &  $\a_2 + \delta_e$  
  & $-\a_2$  & $r_1, r_4$\\\hline
\end{tabular}
\end{center}
\caption{Angle structure; coupling the complex structure moduli to open string modes 
as Fayet-Iliopoulos terms for the
five stack 
SUSY Model II describing the D-brane configuration of table (\ref{spe82}).  
\label{rou1}}          
\end{table}

As the N=1 SUSY models I, II, III  are supersymmetric for the same value of 
complex structure moduli (\ref{li1}) we assume that 
the complex structure moduli departs only slightly from its line of 
marginal stability as
\beq
 U^2 = \frac{\beta_2}{\beta_1}U^3 + \delta_2 \ ,
\eeq
where $\delta_2$ parametrizes the deviation from the supersymmetric 
situation. In this case N=1 supersymmetry will be broken.

For N=1 SUSY Model I, assuming that the departure from the supersymmetric limit is described by table (\ref{tab1}), one finds
\beq
\d_{\a}^{\prime} = \d_{\d}^{\prime} = \delta_a , \  \
\d_{b}^{\prime} = \d_{c}^{\prime} = 0 \ .
\eeq
The sparticle masses for N=1 SUSY model I may be seen in table 
(\ref{seen1})

In table (\ref{rou1}) the deformed angles at the various intersections 
of SUSY Model II are shown. Therefore one may   
find that 
\beq
\delta_a = \frac{3  \rho^2  \beta_1  \delta_2}{[1 + (3  \rho^2  \beta_2  U^3)^2]}, \
\delta_d = \frac{2  \beta_1 \delta_2}{[1 + (2 \beta_2 U^3)^2]},\ 
\delta_e = \frac{\beta_1 \delta_2}{[1 + ( \beta_2 U^3)^2]}, \ \delta_b = \delta_c = 0 \ .
\eeq

The  Fayet-Iliopoulos terms can be expressed in terms of the angle 
deviations from the marginal line. Thus for N=1 Model II, we find
that  
\beq
\xi_a = -\frac{3  \rho^2  \beta_1  \delta_2}{2[1 + (3  \rho^2  \beta_2  U^3)^2]},\  \xi_d = -\frac{ \beta_1 \delta_2}{[1 + (2 \beta_2 U^3)^2]}, \
\xi_e = -  \frac{\beta_1 \delta_2}{2[1 + ( \beta_2 U^3)^2]} \ .
\eeq

\begin{table}[htb] \footnotesize
\renewcommand{\arraystretch}{1.4}
\begin{center}
\begin{tabular}{||c|c|c|c||}
\hline
Sparticle & $(\theta^1,\  \theta^2, \ \theta^3 )$ & Sector& $(mass)^2$ 
\\\hline
 $Q_L$ &  $(-\frac{\pi}{2} - \delta_b^{\prime},\   \a_1 + 
\delta_a^{\prime},\ -\a_1 + \frac{\pi}{2})$ & $(ab)$ & $\frac{1}{2}( \delta_a^{\prime}  - \delta_b^{\prime}   )$  \\\hline
$U_R$ & $(- \frac{\pi}{2} - \delta_c^{\prime},\  \a_1 + \delta_a^{\prime} 
+ \frac{\pi}{2},  
 \ -\a_1  )$ & $(ac)$ & $ \frac{1}{2}(\delta_c^{\prime} -\delta_a^{\prime})      $  \\\hline    
 $D_R$ &   $( \frac{\pi}{2} + \delta_c^{\prime}, \  \a_1 + \delta_a^{\prime} - 
\frac{\pi}{2}, \ \a_1  )$  &  $(ac*)$ &  $ -\frac{1}{2}( \delta_c^{\prime}  
+ \delta_a^{\prime}   )$ \\\hline 
$L$ &   $( -\frac{\pi}{2} - \delta_b^{\prime},\   \a_1 + 
\delta_d^{\prime},\ -\a_1 +
 \frac{\pi}{2} )$  &  $(db) $ & $\frac{1}{2}( \delta_d^{\prime}  - 
\delta_b^{\prime}   )$   
\\\hline    
$N_R$ &   $( -\frac{\pi}{2} - \delta_c^{\prime}, \   \a_1 + 
\delta_d^{\prime} + \frac{\pi}{2},  - \a_1     )$  &  $(dc)$ &  $\frac{1}{2}( \delta_c^{\prime}  - \delta_d^{\prime}   )$   \\\hline    
$E_R$ &   $( \frac{\pi}{2} + \delta_c^{\prime},  \a_1 +  \delta_d^{\prime} 
-  
\frac{\pi}{2},\  -\a_1   )$  &  $(dc*)$ & $-\frac{1}{2}( \delta_c^{\prime}  + 
\delta_d^{\prime}   )$  \\\hline
\end{tabular}
\end{center}
\caption{\small Sparticle masses from Fayet-Iliopoulos terms for 
the four stack quiver of model I seen in figure 2. 
\label{seen1}}
\end{table}

\begin{table}
[htb]\footnotesize
\renewcommand{\arraystretch}{0.8}
\begin{center}
\begin{tabular}{||c||c|c|c|c||}
\hline
\hline
$Brane$ & $\theta_a^1$ & $\theta_a^2$ & $\theta_a^3$ & $approx.\  SUSY \ preserved$\\
\hline\hline
 $a $ & $0$  &
$\a_1 + \delta^{\prime}_a$ & $-\a_1$ & $r_1, r_4$\\
\hline
$b $  & $\frac{\pi}{2} + \delta_b^{\prime}$ & $0$ &
$-\frac{\pi}{2}$  & $ r_2, r_4$ \\
\hline
$c $ & $\frac{\pi}{2} + \delta_c^{\prime}$ &
$-\frac{\pi}{2}$  & $0$ &  $r_3, r_4$\\    
\hline
$d $ & $0$ &  $ \a_1 + \delta_d^{\prime}$  
  & $ -\a_1$  & $r_1, r_4$\\\hline
\end{tabular}
\end{center}
\caption{Angle structure; coupling the complex structure moduli to open string modes as Fayet-Iliopoulos terms for the four stack SUSY Model I.  
\label{tab1}}          
\end{table}

\begin{table}[htb] \footnotesize
\renewcommand{\arraystretch}{1.4}
\begin{center}
\begin{tabular}{||c|c|c|c||}
\hline
Sparticle & $(\theta^1,\  \theta^2, \ \theta^3 )$ & Sector& $(mass)^2$ 
\\\hline
 $Q_L$ &  $(\frac{\pi}{2} + \delta_b,\  - \a_1 - \delta_a,\ \a_1 - \frac{\pi}{2})$ & $(ab)$ & $\frac{1}{2}( \delta_a  - \delta_b   )$  \\\hline
$U_R$ & $( \frac{\pi}{2} + \delta_c,\ - \a_1 - \delta_a - \frac{\pi}{2},  
 \ \a_1  )$ & $(ac)$ & $ \frac{1}{2}(\delta_c -\delta_a)      $  \\\hline    
 $D_R$ &   $(- \frac{\pi}{2} - \delta_c, \ - \a_1 - \delta_a + \frac{\pi}{2}, \ \a_1  )$  &  $(ac*)$ &  $ -\frac{1}{2}( \delta_c  + \delta_a   )$ \\\hline 
$L$ &   $( \frac{\pi}{2} + \delta_b,\  - \gamma_1 - \delta_d,\ \gamma_1 -
 \frac{\pi}{2} )$  &  $(db) $ & $\frac{1}{2}( \delta_d  - \delta_b   )$   \\\hline    
$l_L$ &   $(-\frac{\pi}{2}-  \delta_b,\  \a_2 + \delta_e,\ \frac{\pi}{2}- \a_2   )$  &  $(be) $ & $\frac{1}{2}( \delta_e  - \delta_b   )$   \\\hline    
$N_R$ &   $( \frac{\pi}{2} + \delta_c, \  - \gamma_1 - \delta_d - 
\frac{\pi}{2},  \gamma_1     )$  &  $(dc)$ &  $\frac{1}{2}( \delta_c  - \delta_d   )$   \\\hline    
$E_R$ &   $( -\frac{\pi}{2} - \delta_c, - \gamma_1 -  \delta_d +  
\frac{\pi}{2},\  \gamma_1   )$  &  $(dc*)$ & $-\frac{1}{2}( \delta_c  + 
\delta_d   )$  \\\hline
  $\nu_R$ &   $( -\frac{\pi}{2} - \delta_c,\ \frac{\pi}{2} + \a_2 + \delta_e, 
\ - \a_2)$  &  $(ce)$ &  $\frac{1}{2}( \delta_c  - \delta_e   )$ \\\hline
$e_R$ &   $( -\frac{\pi}{2} - \delta_c,\ \frac{\pi}{2} - \a_2 - \delta_e, \ 
 \a_2    )$  &  $(ce*)$ &  $-\frac{1}{2}( \delta_c  + \delta_e )$ 
 \\\hline
\hline
\end{tabular}
\end{center}
\caption{\small Sparticle masses from Fayet-Iliopoulos terms for 
the five stack quiver 
\label{spe1012}}
\end{table}

{\bf N=1 SUSY Model III}
\newline
For the N=1 SUSY Model III we assume that the departure from the 
supersymmetric 
locus is described in terms of the parameters seen in - appendix A -
table (\ref{ro1}). 
The dependence of the squark masses on the  
Fayet-Iliopoulos terms can be extracted easily and it can be seen 
in table (\ref{spe0}). 
The resulting deviations are :

\beq
{\tilde \delta}_a = {\delta}_a,\ \  {\tilde \delta}_b = {\tilde \delta}_c =
0, \ \ {\tilde \delta}_d = {\tilde \delta}_e = {\tilde \delta}_f = \delta_e
\eeq

Clearly in all the above models the FI terms alone are not 
enough to give to all the squarks 
positive (mass)$^2$, in order to avoid unwanted charge and colour breaking 
minima. The resolution of this puzzle  
 may be found by taking into account loop
effects for the N=1 local SUSY models I, II, III.

\section{Brane recombination}

In this section, we will show that gauge breaking transitions (GBT)
that allow for the breaking of the gauge group without changing
its chiral content exist in models I, II, III. In particular one is 
able to switch between models with
different number of stacks. The end point of the 
recombination process is the - four stack -  N=1 local model 
of table (1), where all U(1)'s beyond 
hypercharge are massive \footnote{due to the Green-Schwarz mechanism}.
That means that the breaking of the 
extra U(1)'s in the five, six stack local N=1 models occurs
at the string scale.

GBT transitions first appeared in brane configurations \cite{D51} with 
 D5-branes wrapping two-cycles in a 
$T^4 \t C/Z_N$ orientifold of type IIB \cite{D5}.
As has been observed in \cite{D51} GBT's also hold for
the toroidal examples of \cite{kokos1, kokos2}. In these models \cite{D51, 
kokos1, kokos2} one is able to show that GBT transitions are able to switch
on - {\em field directions} - between different D-brane stack configurations that 
have the distinct characteristic that they all result in classes of models,
which have only the (non-SUSY) SM at low energy. These 
{\em field directions} cannot be described using only renormalizable 
couplings, thus necessarily their description involves higher order couplings.
For some recent work on the field theoretical description of BR effects
see \cite{BR}.
 
We note that GBT transitions appear as a direct consequence of the
specific way we construct these D6-brane (or D5) configurations. 
In this respect, these transitions, that allow movement between models with different number of stacks, differ from the small
instanton/chirality changing transitions (in the T-dual picture) that 
were considered in
\cite{cve1}. In the latter models, brane recombination (BR) examples
were causing a change in the chiral content among the different BR
phases. We note that BR as an alternative for 
electroweak Higgs breaking mechanism have been 
considered in \cite{iba1}.

\subsection{$Six \ stack \ BR SM's$}

Lets us recombine the branes e, f in the six stack N=1 SUSY Model III. That is we assume that the branes $e, f$ recombine into a new brane ${\tilde e}$,
\beq
e + f = {\tilde e}
\label{er1}
\eeq 

The intersection numbers appearing after brane recombination, may be found
by assuming linearity under homology change, may be seen 
in table (\ref{num1}) and correspond exactly
to the U(1) particle assignments of table (\ref{ska1}).

\begin{table}
\renewcommand{\arraystretch}{1}
\begin{center}
\begin{tabular}{||c|c|c|c|c|c|c|c||}
\hline
Matter Fields & Representation & Intersection & $Q_a$ & $Q_c$ & $Q_d$ & $Q_e$&
\\\hline
 $Q_L$ &  $3(3, 2)$ & $(ab), (ab*)$ & $1$ & $0$ & $0$ & $0$ & $1/6$ \\\hline
$U_R$ & $3({\bar 3}, 1)$ & $(ac)$ & $-1$ & $1$ & $0$ & $0$ & $-2/3$ \\\hline  %
 $D_R$ &   $3({\bar 3}, 1)$  &  $(a c^{\ast})$ &  $-1$ & $-1$ & $0$
 & $0$ & $1/3$ \\\hline 
$L$ &   $2(1, 2)$  &  $({eb}), (eb^{\ast}) $ & $0$ & $0$ & $0$ & $1$ & $-1/2$  \\\hline    
$l_L$ &   $(1, 2)$  &  $(be), (be^{\ast}) $ & $0$ & $0$ & $1$ & $0$ & $-1/2$  \\\hline    
$N_R$ &   $2(1, 1)$  &  $(dc)$ &  $1$ & $0$ & $0$ & $-1$ & $0$  \\\hline    
$E_R$ &   $2(1, 1)$  &  $(dc^{\ast})$ & $-1$ & $0$ & $0$ & $-1$ & $1$  \\\hline
  $\nu_R$ &   $(1, 1)$  &  $(c e)$ &  $0$ & $1$ & $-1$ & $0$ & $0$\\\hline
$e_R$ &   $(1, 1)$  &  $(c e^{\ast})$ &  $0$ & $-1$ & $-1$ & $0$  & $1$ 
\\\hline  
$H_{u,d}$ & $(1,2)$& $(cb*), (cb)$ & $0$   & $ \pm 1 $& $0$ & $0$ & $\mp 1/2$ \\\hline
\hline
\end{tabular}
\end{center}
\caption{\small Fermion spectrum of the five stack 
D6-brane Model-I with its
$U(1)$ charges appearing after the brane recombination (\ref{er1}). 
\label{ska1}}
\end{table}
The reader can be easily convinced that these are exactly the U(1) 
particle intersection numbers
of table (\ref{spe82}), but where the branes d, e have been 
interchanged. Thus no new examples are being generated by this BR.
 Instead, let us try the BR direction;  
recombining the branes d, f; with 
\beq
d + f = {\tilde d}
\label{er2}
\eeq
to see if there are any new examples of vacua left unexplored.
In this case, 
\begin{table}
[htb]\footnotesize
\renewcommand{\arraystretch}{1}
\begin{center}
\begin{tabular}{||c|c|c|c|c|c|c||}
\hline\hline
 $I_{ab}=3$ & $I_{ab*}=3$  & $I_{ac}=-3$ & $I_{ac*}=-3$ & 
$I_{{\tilde e}c} = -2$ &  $I_{{\tilde e}c*} = -2$ & $I_{bc}=-1$ \\\hline

$I_{{\tilde e}b} = 2$ & $I_{{\tilde e}b*} = 2 $ & 
$I_{dc}=-1$  & $I_{dc*}=-1$ &  $I_{db}=1$ & $I_{db*}=1$ & $I_{bc*}=1$\\   
\hline\hline
\end{tabular}
\end{center}
\caption{\small 
Intersection numbers appearing in the five stack model coming after the brane
recombination (\ref{er1}). 
\label{num1}}          
\end{table}
we get exactly the intersection numbers of the five stack N=1 SUSY model
of table (\ref{spe82}). 
Moreover, we further assume the BR combination
\beq
d + e + f = d^{\prime}\ .
\eeq
Our new brane content consists of the $a, b, c, d^{\prime}$. In this case, the chiral
content appearing after BR is that if the four stack SUSY model 
of table (1).

\begin{figure}
\centering
\epsfxsize=5.8in
\hspace*{0in}\vspace*{.0in}
\epsffile{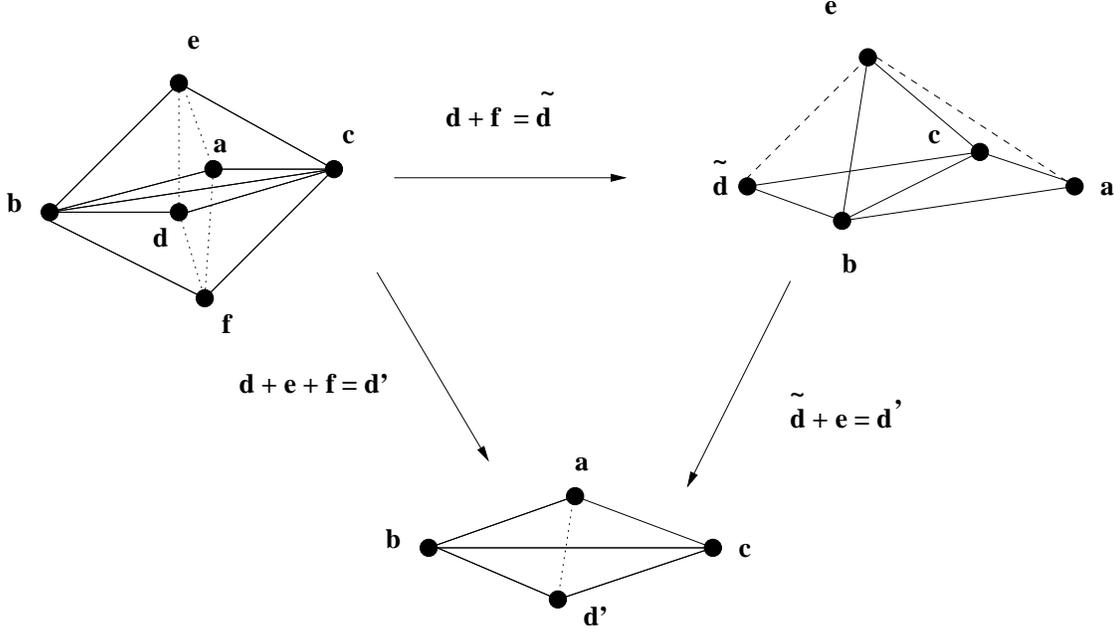}
\caption{\small Brane recombination, homology, flow between six, five and four stack
quivers. Each quiver localizes the spectrum of N=1 Supersymmetric Standard 
Model. The solid lines between the nodes correspond to non-trivial 
intersections. The dotted lines correspond to parallel D6-branes.
}
\end{figure}

\subsection{$Five \ stack \ SM's$}

Lets us now consider the five stack N=1 SUSY Model II and 
recombine the branes d, e. That is we assume that
\beq
d + e = {\tilde d}
\eeq 
Calculating the intersection numbers of the D-brane models surviving the recombination process
we find that the models flow to their four stack counterparts.
In terms of the maximal gauge symmetry in the models in this case, the
following gauge symmetry pattern is realizable \footnote{where we have 
assumed that the brane $e$ has been brought on top of brane $a$.                }
\beq
SU(5) \t SU(2)_L \t SU(2)_R \stackrel{d+e ={\tilde d} }{\longrightarrow} SU(4) \t SU(2)_L \t SU(2)_R
\eeq

The homology flows allowing GBT between Models I, II, III may be
 depicted nicely in terms of quiver diagrams \cite{domo}. Each 
quiver will be depicted in terms of the nodes and their associated 
links. Each node represents the brane content of the brane and its mirror 
image; the link between two nodes representing the non-trivial
intersection between the nodes. For an intersection, respecting N=1 SUSY,
 between two branes a', b', 
the link
will be associated with the presence of 
$I_{ab} (N_{a'}, {\bar N}_{b'}) + I_{ab^{\star}} (N_{a'}, {N}_{b'})$ 
N=1 chiral multiplets. The relevant GBT transitions may be seen in 
figure 2.


\section{Gauge coupling constants}

In this section we will calculate the gauge couplings for the models I, II, 
III. 
The tree level gauge couplings at the string scale are defined 
\footnote{where $\kappa_{\a} =1 $ for $U(N_a)$ and $\kappa_{\a} =2 $ for 
$Sp(2N_a)/SO(2N_a)$ and the gauge fields have been  $tr(T_a T_b) = 1/2$, }
by
\beq
\frac{4\pi}{g^2_{\a}}= \frac{M_{Pl}}{2\sqrt{2} \kappa_{\a} M_s}
\frac{V_{\a}}{\sqrt{V_6}} \ ,
\eeq
where $V_{\a}$ is the volume of the three-cycle that the 
$D6_a$ branes wrap and $V_6$ is the volume of the Calabi-Yau manifold.
A universal property of the models examined in this work, 
is that
the volumes of the three cycles of the $b, c$ branes agree as a 
consequence of the N=1 SUSY condition (\ref{li1}) since 
\beq
\frac{V_b}{V_c}= \frac{\b_2 }{\b_1}\cdot \frac{U^3}{U^2} \stackrel{N=1 \
 SUSY}{=}1 \ .
\eeq

\begin{itemize}\item Four stack models 
\end{itemize}

The four stack model version of  
model I when there is no antisymmetric NS B-field present, that is 
$\b_1= \b_2 =1$, have been proposed in 
\cite{iba2}.  Some consequences for gauge coupling unification for these 
models have been examined in \cite{blulu}.
Let us assume that the gauge group in the four stack model I is that of
$U(3)_c \t SP(2)_b$ ($SU(2)_b$) $\t U(1)_c \t U(1)_d$. Then as the volumes of the 
$a, d$ branes are equal, and the gauge group of the weak SU(2) comes from
an $SP(2)= SU(2)_b$ factor the following relations among the gauge coupling
constants hold
 \beq 
\a_d = \a_a = \a_s, \ \  \a_c = \frac{1}{2}\a_b = \frac{1}{2}\a_w
\eeq
and as a consequence the relation
\beq
\a^{-1}_Y = \frac{2}{3}\frac{1}{\a_s} + \frac{1}{\a_w}
\label{com1}
\eeq
holds at the string scale \cite{blulu}, where \footnote{we have 
introduced a non-trivial NS B-field in the results of \cite{blulu}.} 

\beq
\a^{-1}_Y = \frac{1}{6}\a^{-1}_a + \frac{1}{2}
\a^{-1}_c + \frac{1}{2}\a^{-1}_d 
\eeq

\beqa
\a^{-1}_w &=& \frac{1}{2\sqrt{2}}\cdot
\frac{M_{pl}}{M_s}  \cdot \frac{ \epsilon \sqrt{U^1}}{2 \sqrt{\b_1 \ \b_2}}\ ,           
\label{dio}
\eeqa

\beqa
\a^{-1}_s &=& \frac{1}{2\sqrt{2}}\cdot
\frac{M_{pl}}{M_s}\cdot \left( \frac{1}{\rho^2}\sqrt{\frac{\b_1}{\b_2}
\frac{1}{U_1}}\frac{1}{U_3}
 + 9 \rho^2 ({\epsilon}{\tilde \epsilon})\sqrt{\frac{\b_2}{\b_1} \frac{1}{U_1}} U^3 \right) \ .
\label{ena}
\eeqa
Note that we have assumed that 
the brane $b$ have been placed on top of its orientifold image that is their 
homology classes agree, $\pi_b = \pi_{b^{\star}}$ .
The relation (\ref{com1}) - which is compatible with the   
 SU(5) relation $1/\a_s  = 1/\a_w = (3/5) 1/a_Y$ [ has been derived before 
in \cite{blulu} for the models of \cite{iba2} ] - is not new in the context of 
model building in string compactifications as it has been also derived     
in the context of type IIB 4D orientifold compactifications \cite{rigo}.
We can further assume that the initial gauge symmetry in the models is 
$U(4) \t SU(2)_L^b \t SU(2)_R^c$, that is the brane $d$ has been placed on top
of brane $a$, and the homology classes of $b,c$ branes agree 
($V_b \rightarrow V_c$). It follows  
 \beq 
\a_d = \a_a = \a_s, \ \  \a_c = \a_b = \a_w
\label{com0}
\eeq
and as a consequence the relation
\beq
\a^{-1}_Y = \frac{2}{3} \a^{-1}_s + \frac{1}{2} \a^{-1}_w
\label{com2}
\eeq
holds at the string scale.

\begin{itemize}\item Five stack models 
\end{itemize}
Assuming that the initial gauge symmetry at the string scale is 
given by (\ref{assu1}) the following relations among the gauge couplings hold 
\beq
\a^{-1}_Y = \frac{1}{6}\a^{-1}_a + \frac{1}{2}
\a^{-1}_c + \frac{1}{2}\a^{-1}_d + \frac{1}{2}\a^{-1}_e  
\eeq
\beqa
\a^{-1}_d &=& \frac{1}{2\sqrt{2}}\cdot
\frac{M_{pl}}{M_s}\cdot \left( \sqrt{\frac{\b_1}{\b_2}
\frac{1}{U_1}}\frac{1}{U_3}
 + 4 ({\epsilon}{\tilde \epsilon})\b_1 \b_2 \sqrt{\frac{\b_2}{\b_1} \frac{1}{U_1}} U^3 \right)  ,\\
\a^{-1}_e &=& \frac{1}{2\sqrt{2}}\cdot
\frac{M_{pl}}{M_s}\cdot \left( \sqrt{\frac{\b_1}{\b_2}
\frac{1}{U_1}}\frac{1}{U_3}
 + ({\epsilon}{\tilde \epsilon}) \b_1 \b_2 \sqrt{\frac{\b_2}{\b_1} \frac{1}{U_1}} U^3 \right)
\label{diot1}
\eeqa
\beq
 \ \a_c =  \frac{1}{2}\a_b  =   \frac{1}{2}\a_w, \ \ 
\eeq
where ${\ti \a}_w$, ${\ti \a}_a$ are given in (\ref{dio}), (\ref{ena}) respectively.

\begin{itemize}\item Six stack models 
\end{itemize}
Let us assume that the initial gauge symmetry in the models is as in
(\ref{da12}).  Then the gauge couplings are as follows:   
\beq
{\tilde \a}^{-1}_Y = \frac{1}{6}{\tilde \a}^{-1}_a + \frac{1}{2}
{\tilde \a}^{-1}_c + \frac{1}{2}{\tilde \a}^{-1}_d + \frac{1}{2}
{\tilde \a}^{-1}_e  +  
\frac{1}{2}{\tilde \a}^{-1}_f \ ,
\eeq
\beq
{\tilde \a}_c = \frac{1}{2}\a_b =  \frac{1}{2}\a_w, \ 
 \ \ {\tilde \a}^{-1}_d =  {\tilde \a}^{-1}_e = {\tilde \a}^{-1}_f\ ,
\eeq
\beq
a_d^{-1} = \frac{1}{2 \sqrt{2}}\frac{M_{pl}}{M_s}
\left( \sqrt{\frac{1}{U_1 U_2 U_3}} + {\epsilon}{\tilde \epsilon}
{ \beta}_1 {\beta}_2 \sqrt{\frac{U^2 U^3}{U^1}}      \right)
\eeq
and $\a_w$, $\a_a$ are given in (\ref{dio}), (\ref{ena}) respectively.

As the D6-brane configurations of tables (1), (4), (7)  
localize the spectrum of N=1 MSSM in cases where $\b_1 = \b_2 = 1$, the 
five and the six stack models they may be expected to exhibit 
 \footnote{As it has also 
been exhibited for model I, when $\b_1 = \b_2 = 1$,
in \cite{blulu}.}
unification of the gauge coupling 
constants at a scale of the order of $10^{16}$ GeV and for 
appropriate values of the complex moduli.  This is possible as
the models possess enough freedom - through their dependence on the 
string scale and the complex structure moduli. 

\section{Gauge mediated N=1 SUSY breaking}

The models we have constructed in the present work do not satisfy the 
RR tadpoles (\ref{tadpo}). However as the gauge and mixed U(1) anomalies 
for the
D-brane configurations of tables (1), (4), (7) cancel, 
RR tadpoles may be
cancelled by adding a non-factorizable D6 brane (NF).
As the presence of the NF brane is not expected to create fields  
on intersections that will survive massless to low energies
\footnote{as it has been noticed in the 
first reference of \cite{iba1}.},  
no new open string 'visible' sectors may be 
created but massive fields.  
Also as ${\em Str}M^2$ in a supersymmetric 
theory is expected to vanish one may expect that 
SUSY breaking is generated beyond tree level. In this case the couplings 
responsible for breaking N=1 SYSY may be non-renormalizable.
As a consequence of this remark, this phase of the theory
is similar to the one we have found in the Pati-Salam models - that possess
N=1 supersymmetric subsectors - 
of \cite{kokos3}, where the  
presence of extra branes was breaking the extra U(1)'s surviving the 
Green-Schwarz mechanism. In the latter case the extra branes create 
gauge singlets that make 
massive through non-renormalizable couplings all exotic fermions that 
transform under both the extra -plays the role of an 'NF' - brane and the visible SM branes; the latter fields playing the role of messenger fields.  
\newline       
The NF D6-brane in the present constructions is
 expected to break the N=1 SUSY respected by the D6-branes on 
intersections and thus 
plays the role of a supersymmetry breaking (SB) sector. This SB sector 
may be expected to break N=1 
supersymmetry by gauge mediated supersymmetry breaking (GMSB)[see for 
example \cite{dine}.] 
The NF brane, as the 4D models we examine are toroidal orientifolds of type 
IIA, may be expected, in the general case, to preserve 
$N = 0$ SUSY. 
 In this case the one-loop contribution
to gauginos may be depicted as in figure (3). As usual in gauge mediation, the
mass of the gauginos may be of order 
\beq
m_{\lambda_i} \approx c_i \frac{\a_i}{4 \pi}M_s \ ,
\eeq 
where we have assumed that the supersymmetry breaking scale is at the string 
scale. As one-loop contributions
to squarks and sleptons may vanish 
Higgs masses will get corrections from two loop effects \footnote{see relevant
comments in the first 
reference of \cite{iba1}.} of the order $\sim (\a/4 \pi)M_s$, which limits 
the scale of electroweak symmetry breaking at small values as long as 
$M_s \leq 30$ TeV. Precise bounds on $M_s$ will be derived once the NF brane
is available.    
The models constructed in this work that localize the spectrum of the N=1 
SM use intersecting D6-branes. One can imagine a scenario of this sort in the 
presence of a mixture of orthogonal 
 D3, D7 branes, as in the type I compactifications described in \cite{rigo}, 
where the gauge group of the N=1 
SM may originate from the D3$_a$D3$_a$ sector and the role of the
NF brane in these models is played by the  D7$_a$D7$_a$ sector. In this case,
supersymmetry breaking will be transmitted to the visible N=1 SM sector 
by the gauge interactions of the massive fields from the 
D3$_a$D7$_b$ sector.

\begin{figure}
\centering
\epsfxsize=6in
\hspace*{0in}\vspace*{.0in}
\epsffile{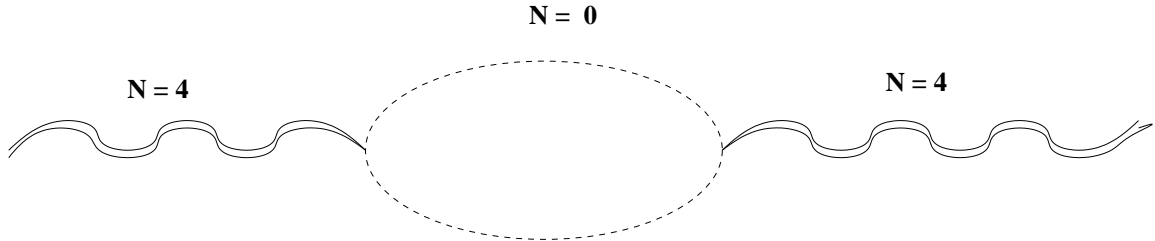}
\caption{\small One loop contributions to gaugino masses from heavy states, belonging
to the NF brane, running
in the loop. The possible supersymmetries respected by the NF brane are 
shown. 
}
\end{figure}

However, the  
presence of this 
NF brane is rather out of reach at present \footnote{we hope to refer to these issues in 
the future}. As a consequence of this difficulty we have chosen to 
break N=1 SUSY from the start by using the presence of 
Fayet-Iliopoulos terms. We have found that at tree level it is not 
possible to give positive (mass)$^2$ to all sparticles \footnote{The same observation
was made for the 'triangular' quivers of \cite{iba1} that possess the same
N=1 properties as the present constructions}. By
calculating the two-loop corrections to these masses, in order to guarantee
the presence of positive (mass)$^2$ for all sparticle masses,
we may be able to avoid the existence of charge and colour breaking minima.
We note that it is also possible that the existence of FI-terms 
triggers the generation of  \footnote{We thank Angel Uranga for 
reminding us of this possibility.} vev's for the some scalars
which restabilizes the vacuum and restores the local N=1 susy on the visible
branes. These issues may be clarified when also two loop  
corrections may be 
calculated.  
We hope 
to return to these issues in the future.

\subsection{Conclusions}

In this work, we have constructed four dimensional string theory models, 
which in some cases, localize
the MSSM spectrum when the NS B-field is not present across
 the six dimensional internal toroidal space. 
We have also constructed
extensions of the MSSM with the same fermion and boson spectrum but where
we have present extra - beyond the standard MSSM $H_u$, $H_d$ 
 - Higgs multiplets (HM's). The number of HM's depend on the 
existence of the NS B-field $ \#_{Higgs} =1/(\b_1 \b_2)$ \footnote{where 
 $\{\b_1 , \b_2 \}= \{1, 1/2\}$ 
the value of the B-field in the second and third T$^2$. The fractional
values correspond to the presence of a non-trivial B-field.}.

The most important property shared by these constructions is that all visible
intersections, of the D-brane configurations of tables (1), (4), (7) 
share the same N=1 SUSY. Once the supersymmetry 
breaking sector is found -  the explicit form of the NF brane - we will 
be able to show explicitly the breaking of N=1 supersymmetry by gauge 
mediation.
 
As we have described in section six we are able to move 
between
models with different number of stacks. This is achieved 
using brane recombination (BR). BR describes the homology flow 
deformations of models II, III around the four stack model I. 
The presence of this mechanism 
is compatible with the deformation of the flat directions in the line of marginal stability of the 
D-term potential. To follow the present BR directions we have to deform
along the lines of marginal stability.  
We note that even though there is a homology flow 
between the different models, they are not 
equivalent at the quantum level. The Yukawa interactions of models I, II, II 
are different. In addition, unless the full effective potential is 
calculated we cannot be sure that a specific BR direction is preferred 
against another one.

Baryon and lepton number are gauged symmetries in the present models as the 
corresponding gauge bosons receive a mass through the generalized 
Green-Schwarz mechanism.
Thus proton is stable and only Dirac mass terms are allowed for neutrinos.
Using Model II (five stack) against Model I - or Model III
against Model II and/or Model I - for the explanation of charged lepton and 
neutrino mass mixing is certainly more advantageous as 
localizing each charged lepton and neutrino to different intersections
allows for more freedom to generate the quark and lepton hierarchies. 
For example neutrino hierarchies may most
easily accommodated in the five stack model II, as two generations of 
neutrinos get localized at the 
intersection $cd$; the third generation localized at the intersection $ce$. 
In this case two generations of neutrinos may be shown to 
be heavier 
than the third one, 
as the SuperKamiokande results suggest \footnote{By construction in 
the present model II, as in the five stack 
D6-models of \cite{kokos1}, two neutrino and charged lepton species 
get localized at different intersection points.}. It will be 
interesting to discuss extensively the flavour problem, the origin of 
families and 
fermion - quarks and lepton - masses as the latter appears to be consistent 
with the 
implementation of 'texture zeros' \cite{ro} that points towards 
beyond the SM symmetries.

Finally, we examined the modification of the gauge coupling constants
in the presence of a NS B-field. We obtained several interesting relations 
among the gauge coupling constants at the string scale that may 
be used to examine the consequences for the gauge coupling running 
unification (GCU) of the MSSM models I, II and III 
(and their extensions in the presence of a non-zero NS B-field; its net effect
the addition of extra Higgs MSSM multiplets with identical quantum numbers). 
 The GCU in this case should
appear at a relative low scale and below 30 TeV 
as a consequence of gauge mediation and the existence of the SUSY
breaking sector.  
 This is to be contrasted
with the `observed' unification of gauge couplings at the gauge theory 
version of the
 MSSM at a GUT scale of $\approx 10^{16}$ GeV and the GCU in models of 
4D weak coupling N=1 heterotic orbifold compactifications \cite{nano}
at $\approx 10^{17}$ GeV. The gauge coupling dependence of the present models 
on the complex moduli
values allows enough freedom to possibly accommodate such unification 
choices.   
These issues will be examined in detail 
elsewhere.
 
\begin{center}
{\bf Acknowledgments}
\end{center}
I am  grateful
to L.~E.~Ib\'a\~nez, D. Cremades, A. Uranga
for useful discussions. 
I would like also to thank the Institute of
Particle Physics Phenomenology at Durham - and the Organizers of 
the 2nd String Phenomenology Conference -
for their warm hospitality,
 where part of this work was done.


\section{Appendix A}
In this appendix further details for Model III may be found.
\begin{table}
[htb]\footnotesize
\renewcommand{\arraystretch}{1}
\begin{center}
\begin{tabular}{||c||c|c|c|c||}
\hline
\hline
$Brane$ & $\theta_a^1$ & $\theta_a^2$ & $\theta_a^3$ & $approx. \ SUSY \ preserved$\\
\hline\hline
 $a $ & $0$  &
$\a_1 + {\tilde \delta}_a$ & $-\a_1$ & $r_1, r_4$\\
\hline
$b $  & $\frac{\pi}{2} + {\tilde \delta}_b$ & $0$ &
$-\frac{\pi}{2}$  & $ r_2, r_4$ \\
\hline
$c $ & $\frac{\pi}{2} + {\tilde \delta}_c$ &
$-\frac{\pi}{2}$  & $0$ &  $r_3, r_4$\\    
\hline
$d $ & $0$ &  $ \a_2 + {\tilde \delta}_d$  
  & $ -\a_2$  & $r_1, r_4$\\\hline
$e $ & $0$ &  $\a_2 + {\tilde \delta}_e$  
  & $-\a_2$  & $r_1, r_4$\\\hline
$f $ & $0$ &  $\a_2 + {\tilde \delta}_f$  
  & $-\a_2$  & $r_1, r_4$\\\hline
\end{tabular}
\end{center}
\caption{Angle structure, coupling the complex structure moduli to open string modes 
as Fayet-Iliopoulos terms for the
six stack 
SUSY Model III.  
\label{ro1}}         
\end{table}
These two tables are  
 related to the 
angle structure and the sparticle masses coming after the introduction of the
Fayet-Iliopoulos terms.
\begin{table}
[htb] \footnotesize
\renewcommand{\arraystretch}{1}
\begin{center}
\begin{tabular}{||c|c|c|c||}
\hline
Sparticle & $(\theta^1, \ \theta^2, \ \theta^3 )$ & Sector& $(mass)^2$ 
\\\hline
 $Q_L$ &  $(- \ \frac{\pi}{2} - \  {\tilde \delta}_b,\   \  \a_1 +  
{\tilde \delta}_a,\ - \a_1 + \frac{\pi}{2})$ & $(ab)$ & $\frac{1}{2}( \delta_a  - \delta_b   )$  \\\hline
$U_R$ & $( - \ \frac{\pi}{2} -  {\tilde \delta}_c,\  \a_1 +  
{\tilde \delta}_a + \frac{\pi}{2},  
 \ - \  \a_1  )$ & $(ac)$ & $ \frac{1}{2}( \delta_a  - \delta_c   )   $  \\\hline    
 $D_R$ &   $( \frac{\pi}{2} +  {\tilde \delta}_c, \  \a_1 + 
 {\tilde \delta}_a - \frac{\pi}{2}, \ - \ \a_1  )$  &  $(ac*)$ &  $ -\frac{1}{2}( \delta_c  + \delta_a   )$ \\\hline 
$L^1$ &   $( - \frac{\pi}{2} -  {\tilde \delta}_b,\  \a_2 +  
{\tilde \delta}_d,\ - \ \a_2 +
 \frac{\pi}{2} )$  &  $(db) $ & $\frac{1}{2}( \delta_d  - \delta_b   )$   \\\hline   
$L^2$ &   $(\frac{\pi}{2} +  {\ti \delta}_b,\  - a_2 - {\ti \delta}_e,\ 
- \ \frac{\pi}{2} + \a_2   )$  &  $(be) $ & $\frac{1}{2}( \delta_e  - \delta_b   )$   \\\hline 
 $L^3$ &   $(\frac{\pi}{2} +  {\ti \delta}_b,\ - a_2 - {\ti \delta}_f,\ -
\frac{\pi}{2} + \a_2   )$  &  $(bf) $ & $\frac{1}{2}( \delta_f  - \delta_b   )$   \\\hline   
$N_R^1$ &   $( \frac{\pi}{2} + {\ti \delta}_c, \  - \a_2 - {\ti \delta}_d - 
\frac{\pi}{2}, \  \a_2     )$  &  $(cd)$ &  $\frac{1}{2}( \delta_c  - \delta_d   )$   \\\hline    
$E_R^1$ &   $( \frac{\pi}{2} + {\ti \delta}_c, \ \a_2 +  {\ti \delta}_d -  
\frac{\pi}{2},\  - \  \a_2   )$  &  $(cd*)$ & $\frac{1}{2}( \delta_c  + 
\delta_d   )$  \\\hline
$N_R^2$ &   $( \frac{\pi}{2} + {\ti \delta}_c, \  - \a_2 - {\ti \delta}_e - 
\frac{\pi}{2},  \ \a_2     )$  &  $(ce)$ &  $\frac{1}{2}( \delta_c  - 
\delta_e   )$   \\\hline    
$E_R^2$ &   $( \frac{\pi}{2} + {\ti \delta}_c, \  \a_2 +  {\ti \delta}_e -  
\frac{\pi}{2},\  - \ \a_2   )$  &  $(ce*)$ & $-\frac{1}{2}( \delta_c  + 
\delta_e   )$  \\\hline
$N_R^3$ &   $( \frac{\pi}{2} + {\ti \delta}_c, \  - \a_2 - {\ti \delta}_f - 
\frac{\pi}{2}, \  \a_2     )$  &  $(cf)$ &  $\frac{1}{2}( \delta_c  - \delta_f   )$   \\\hline    
$E_R^3$ &   $( \frac{\pi}{2} + {\ti \delta}_c, \  \a_2 +  {\ti \delta}_f -  
\frac{\pi}{2},\  - \a_2   )$  &  $(cf*)$ & $-\frac{1}{2}( \delta_c  + 
\delta_f   )$  
 \\\hline
\end{tabular}
\end{center}
\caption{\small Sparticle masses from Fayet-Iliopoulos terms for the
six stack quiver 
\label{spe0}}
\end{table}
We note that in table (\ref{spe0}) we have assumed that $|\frac{\pi}{2} 
- \a_1| > 0$, $|\frac{\pi}{2} 
- \a_1| > |\delta_a  - \delta_b|$, are satisfied. Also we assume that the relations 
$|\frac{\pi}{2} 
- \a_2| > 0$, 
$|\frac{\pi}{2} 
- \a_2| > |\delta_d  - \delta_b|$, $|\frac{\pi}{2} 
- \a_2| > |\frac{\delta_a}{2}|$, $|\frac{\pi}{2} 
- \a_2| > |\delta_e  - \delta_b|$ hold.


\end{document}